\documentclass[trackchanges, twocolumn]{aastex7}

\newcommand\aastex{AAS\TeX}
\newcommand{\galaxylong}{SMILES-GS-191748}
\newcommand{\galaxy}{Eridu}
\newcommand{\OIII}{[O~{\rm \scriptsize III}]$\lambda\lambda4960,5008$}
\newcommand{\NII}{[N~{\rm \scriptsize II}]$\lambda\lambda6550,6585$}
\newcommand{\SII}{[S~{\rm \scriptsize II}]$\lambda\lambda6718,6732$}
\newcommand{\HeII}{He~{\rm \scriptsize II}$\lambda4686$}

\begin{document}

\title{Excavating The Ruins: an Ancient $z=2.675$ Galaxy Which Formed in the First 500 Myr}

\author[orcid=0000-0002-2446-8770,sname='McConachie']{Ian McConachie}
\affiliation{Department of Astronomy, University of Wisconsin-Madison, 475 N. Charter St., Madison, WI 53706 USA}
\email[show]{ian.mcconachie@wisc.edu}

\author[0000-0002-0243-6575]{Jacqueline Antwi-Danso}
\altaffiliation{Banting Postdoctoral Fellow}
\affiliation{David A. Dunlap Department of Astronomy and Astrophysics, University of Toronto, 50 St. George Street, Toronto, Ontario, M5S 3H4, Canada}
\email[]{j.antwidanso@utoronto.ca}

\author[0000-0003-2144-2943]{Wenjun Chang}
\affiliation{Department of Physics and Astronomy, University of California, Riverside, 900 University Avenue, Riverside, CA 92521, USA}
\email[]{wenjun.chang@email.ucr.edu}

\author[0000-0003-1371-6019]{M. C. Cooper}
\affiliation{Center for Cosmology, Department of Physics and Astronomy, University of California, Irvine, Irvine, CA, USA}
\email[]{cooper@uci.edu}

\author[]{Adit Edward}
\affiliation{Department of Physics and Astronomy, York University, 4700, Keele Street, Toronto, ON MJ3 1P3, Canada}
\email[]{ahedward@yorku.ca}

\author[0000-0001-6003-0541]{Ben Forrest}
\affiliation{Department of Physics and Astronomy, University of California, Davis, One Shields Avenue, Davis, CA 95616, USA}
\email[]{bforrest@ucdavis.edu}

\author{Percy Gomez}
\affiliation{W.M. Keck Observatory, 65-1120 Mamalahoa Hwy., Kamuela, HI 96743, USA}
\email[]{pgomez@keck.hawaii.edu}

\author[0000-0002-9158-6996]{Han Lei}
\affiliation{Department of Physics, McGill Space Institute, McGill University, 3600 rue University, Montr\'{e}al, Qu\'{e}bec H3A 2T8, Canada}
\email[]{han.lei@mail.mcgill.ca}

\author{Zach J. Lewis}
\affiliation{Department of Astronomy, University of Wisconsin-Madison, 475 N. Charter St., Madison, WI 53706 USA}
\email[]{zjlewis@wisc.edu}

\author[0000-0001-9002-3502]{Danilo Marchesini}
\affiliation{Department of Physics \& Astronomy, Tufts University, MA 02155, USA}
\email[]{Danilo.Marchesini@tufts.edu}

\author[0000-0003-0695-4414]{Michael V. Maseda}
\affiliation{Department of Astronomy, University of Wisconsin-Madison, 475 N. Charter St., Madison, WI 53706 USA}
\email[]{maseda@astro.wisc.edu}

\author[0000-0002-9330-9108]{Adam Muzzin}
\affiliation{Department of Physics and Astronomy, York University, 4700, Keele Street, Toronto, ON MJ3 1P3, Canada}
\email[]{muzzin@yorku.ca}

\author{Allison Noble}
\affiliation{School of Earth and Space Exploration, Arizona State University, Tempe, AZ 85287, USA}
\affiliation{Beus Center for Cosmic Foundations, Arizona State University, Tempe, AZ 85287 USA}
\email[]{allisongnoble@gmail.com}

\author[0000-0001-8169-7249]{Stephanie M. Urbano Stawinski}
\affiliation{Center for Cosmology, Department of Physics and Astronomy, University of California, Irvine, Irvine, CA, USA}
\affiliation{Department of Physics, University of California, Santa Barbara, Santa Barbara, CA 93106, USA.}
\email[]{sstawins@uci.edu}

\author{Tracy Webb}
\affiliation{Department of Physics, McGill Space Institute, McGill University, 3600 rue University, Montr\'{e}al, Qu\'{e}bec H3A 2T8, Canada}
\email[]{webb@physics.mcgill.ca}

\author[0000-0002-6572-7089]{Gillian Wilson}
\affiliation{Department of Physics, University of California, Merced, 5200 Lake Road, Merced, CA 95343, USA}
\email[]{gwilson@ucmerced.edu}

\author[0000-0002-6505-9981]{M.E. Wisz}
\affiliation{Department of Physics, University of California, Merced, 5200 Lake Road, Merced, CA 95343, USA}
\email[]{mwisz@ucmerced.edu}

\begin{abstract}

We present the analysis of an ancient galaxy at $z=2.675$ which we dub ``\galaxy.'' Simultaneously modeling the JWST/NIRSpec G140M and G235M spectra from the SMILES program and $0.4-25\ \mu\mathrm{m}$ HST, JWST/NIRCam, and JWST/MIRI photometry from the the JADES+SMILES photometric catalogs shows that \galaxy\ is massive and quiescent with stellar mass $\log(M_*/\mathrm{M_\odot})=10.96^{+0.01}_{-0.01}$ and average star formation rate $<1\ \mathrm{M_\odot\ yr^{-1}}$ over the last 100 Myr. Star formation histories inferred from various models produce disconcertingly early and fast formation within $\sim300$ Myr of the Big Bang and quenching 2 Gyr prior to observation ($z\sim10$). This stellar mass assembly implies that the progenitor of \galaxy\ had $M_*\approx10^{11}\ \mathrm{M_\odot}$ at $z>10$, nearly two orders of magnitude more than the most massive current high redshift observations. From \galaxy's spectrum we infer $\mathrm{[Mg/Fe]} =+0.65^{+0.20}_{-0.19}$, indicating its stellar population is extremely $\alpha$-enhanced, which is consistent with the rapid formation timescale inferred from its star formation history. \galaxy\ inhabits a massive protostructure which offers additional explanations for rapid mass assembly and quenching via environmental mechanisms, e.g. major mergers. Though its inferred formation is at odds with observations of the brightest cosmic dawn galaxies, we anticipate that future high-redshift galaxy formation models and sophisticated stellar population modeling codes will unearth how \galaxy\ formed at the dawn of time.

\end{abstract}

\keywords{\uat{Galaxies}{573} --- \uat{Galaxy evolution}{594} --- \uat{Quenched Galaxies}{2016} --- \uat{High-redshift galaxies}{734}}

\section{Introduction}

Observations of massive and quiescent galaxies have historically been used as strong tests on models for cosmology and galaxy formation and evolution \citep[e.g., ][]{Dunlop1996, Cimatti2004}. Confirmations of quiescent galaxies at increasingly high redshift and stellar masses challenged simulations to efficiently build stars and quench quickly enough to reproduce observations \citep[][]{Somerville2015, Beckmann2017a, Donnari2020, Donnari2021, DeLucia2024, Lagos2024}. However, in the pre-JWST era, studies of quiescent galaxies were limited to all but the brightest and most massive (with $\log M_*/M_{\odot} \gtrsim 11$) at $z\lesssim4$. Still, a remarkable number of old and massive galaxies have been observed and confirmed from the ground \citep[e.g., ][]{Kriek2016a, Newman2015, Newman2018, Belli2019, Glazebrook2017, Schreiber2018a, Schreiber2018b, Forrest2020a, Forrest2020,Kalita2021, Kubo2021b, Ito2023a, Tanaka2024a, Kakimoto2024a, Antwi-Danso2025}. Many of these confirmed galaxies had recent quenching times ($3\lesssim z\lesssim 5$), as the stellar populations of theoretical older or higher redshift quiescent galaxies would be fainter and harder to confirm with existing observatories.

The launch of JWST has granted astronomers unparalleled access to study galaxy formation and evolution in the early universe. Besides spectroscopically confirming galaxies out to $z>10$ \citep{Bunker2023, Castellano2024, Tacchella2023, ArrabalHaro2023, Curtis-Lake2023, Carniani2024, Carniani2025, Naidu2025}, it has also enabled the confirmation of quiescent galaxies at redshifts hitherto inaccessible by other ground- and space-based observatories \citep[]{Carnall2023a, Carnall2024, DeGraaff2024a, Weibel2025, Onoue2024}. The sensitivity and wavelength coverage of NIRCam, MIRI, and NIRSpec have enabled detailed investigations of the populations of quiescent galaxies at signal-to-noise (S/N) levels that were previously infeasible as well, allowing for study of their number densities, stellar populations, and other properties (e.g., \citealt{Park2024a, Slob2024, Beverage2025, Nanayakkara2024, Nanayakkara2025, Baker2025, Kawinwanichakij2025, Ito2025b}; Forrest et al. 2025, in prep.). Many of these recent studies have found that their subjects formed relatively recently prior to observation with formation timescales that are faster than what is seen in the low redshift universe, as was typical for pre-JWST studies.

However, the ancient $z=3.2$ quiescent galaxy with a PRISM spectrum in \citet{Glazebrook2024} appeared so old and so massive that its existence seemed to challenge $\Lambda$CDM cosmology \citep{Behroozi2018, Boylan-Kolchin2023}. ZF-UDS-7329 was first proposed to have formed at $z\approx11$, by which time dark matter halos would not have been massive enough to host such a galaxy. Analysis of high S/N medium resolution NIRSpec grating spectra from the EXCELS program \citep{Carnall2024} confirmed that ZF-UDS-7329 did indeed form extremely early, but the star formation history (SFH) inferred from the fit to medium resolution spectra depicted a mass assembly more in-line with cosmology (though a near-maximum baryon conversion efficiency would be required). \citet{Carnall2024} also found that ZF-UDS-7329 quenched at $z\approx6$ and was $\alpha$-enhanced. Further analysis of the PRISM spectrum with a variety of models by \citet{Turner2025} confirmed the earlier findings of ZF-UDS-7329's age and that it formed very efficiently, but that its existence is not in direct conflict with cosmology.

$\Lambda$CDM cosmology has withstood assaults from other avenues as well. Spectroscopic followup of over-massive high-redshift candidates initially identified in photometry \citep{Labb2023} revealed that these ``Universebreakers'' are not and do not. Instead they comprise a new class of interesting objects called ``Little Red Dots'' (e.g, \citealt{Matthee2024, Wang2024, Perez-Gonzalez2024, Kokorev2024, Yue2024, Taylor2025}), the light of which is unlikely solely stellar in origin. High-redshift quiescent galaxies observed with the NIRSpec PRISM suggest early formation times which \emph{could} conflict with cosmological models, but the low resolution of the PRISM and difficulty modeling the spectral energy distributions (SEDs) of fast-forming (likely $\alpha$-enhanced) stellar populations has prevented strong conclusions from being drawn \citep{DeGraaff2024a, Weibel2025}. Bright, extremely high-redshift galaxy candidates \citep[e.g.,][]{Naidu2022} have been identified, but follow-up spectroscopy has revealed them to be lower-redshift galaxies with high dust obscuration \citep[e.g.,][]{ArrabalHaro2023}. Thus far, $\Lambda$CDM seems robust to observations of high-redshift galaxies.

Still, JWST has enabled the detection, confirmation, and characterization of a remarkable population of $z>10$ galaxies at ``cosmic dawn'' \citep{Castellano2024, Tacchella2023, ArrabalHaro2023, Curtis-Lake2023, Carniani2024, Naidu2025}. These galaxies are considered high mass for their epoch ($\log\ M_*/M_\odot\sim8-9$), are forming stars at rates of SFR$\ \sim1-30\ \rm M_\odot\ yr^{-1}$, and are remarkably luminous. While relatively diminutive compared to the monster galaxies of the low-redshift universe ($z<5$), these galaxies represent the ``upper limits'' of formation allowed by cosmology and are pushing the limits of galaxy formation models \citep[e.g.,][]{Dekel2023, Sun2023a, Shen2024}. Finally, it is tautologically clear that since these galaxies existed at cosmic dawn, their descendants should be visible at cosmic noon.

In this work we present \galaxylong\ (nicknamed ``\galaxy''), an ancient and massive quiescent galaxy at $z=2.675$, the inferred early and rapid formation of which is in tension with even the highest redshift observations. The paper is organized as follows: in \S\ref{sec:obs} we present the imaging and spectroscopy of \galaxy. We fit the photometry and spectroscopy with \texttt{Prospector} to infer stellar population parameters and SFHs in \S\ref{sec:modeling}, and we discuss how our recovered SFHs compare with other massive galaxies and $z>10$ observations. We check our modeling by also fitting \galaxy\ with \texttt{Bagpipes} in \S\ref{sec:pipes} and discuss the impact of the assumed SFH parameterization. In \S\ref{sec:alfalpha}, we measure stellar abundances from prominent absorption features in the spectrum using \texttt{alf$\alpha$} and discuss its $\alpha$-enhancement. We consider \galaxy's environment and its role in \S\ref{sec:env}, and we summarize our main conclusions in \S\ref{sec:conc}. In this work, we assume best-fit cosmological parameters from WMAP 9 year results \citep{Hinshaw2013}: $H_0 = 69.32 \ \mathrm{km \ s^{-1} \ Mpc^{-1}}$, $\Omega_m = 0.2865$, and $\Omega_\Lambda = 0.7135$ and utilize a Chabrier initial mass function \citep{Chabrier2003a} (unless stated otherwise, as required by certain SED modeling codes). All magnitudes are on the AB system \citep{Oke1983}. All logarithms are base 10 unless stated otherwise.

\section{\galaxylong\ and Observed Data} \label{sec:obs}

\begin{figure*}[!htb]
\includegraphics[width=\linewidth]{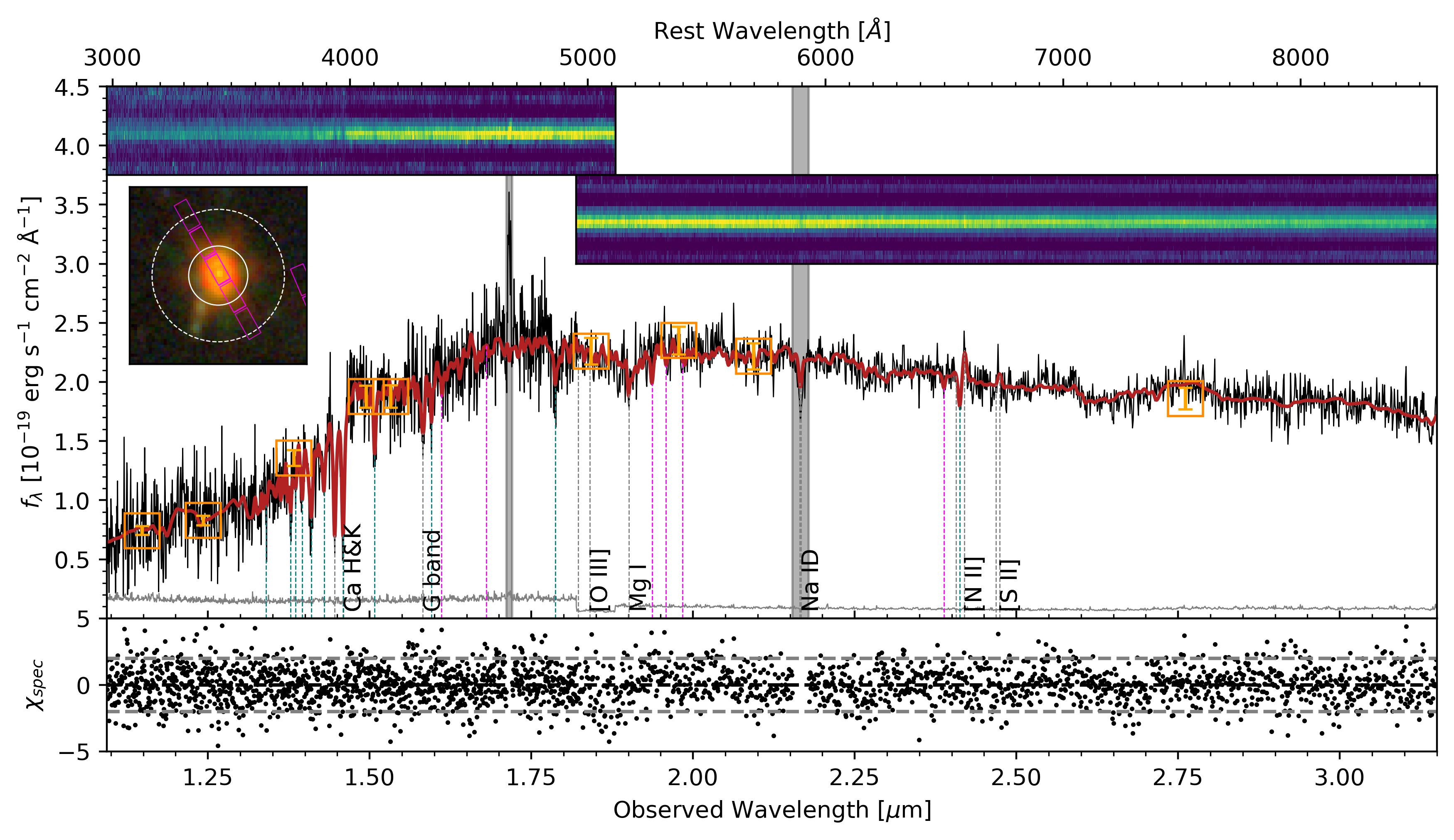}
\caption{The observed combined spectrum (black), associated uncertainty (gray, along the bottom), and best-fit \texttt{Prospector} model spectrum (red). Grayed out regions (cosmic ray contamination, Na ID absorption) are masked out when modeling the spectrum. The locations of features are indicated with dashed lines; teal corresponds to hydrogen (Balmer), pink indicates to iron, and gray is other metals (only some of which are labeled; for all labeled lines see Figure \ref{fig:spec_zoom}). Observed photometry is shown as orange open squares (for the full model posterior in the $0.4-25\ \mu$m range, see Figure \ref{fig:phot_model}). At the top we show the 2D spectra for the G140M and G235M grating spectra. Our Prospector model uses a 10-degree Chebyshev polynomial to scale the spectrum to the model while simultaneously fitting both the spectrum and the photometry. The scaling polynomial is applied to the spectrum in this figure to illustrate the good match between the model, spectrum, and observed photometry (a zoom-in on the unscaled spectrum is shown in Figure \ref{fig:spec_zoom}). We show the residuals for the spectroscopy (black points) along the bottom of the figure (residuals from photometry are shown in Figure~\ref{fig:phot_model}). In the inset on the left we show an RGB F115W/F277W/F444W color cutout of \galaxy\ with the MSA shutter placement outlined in magenta, the $r=0.25^{\prime\prime}$ aperture as a solid white line, and a $r=0.56^{\prime\prime}$ radius circle as a dashed white line. The ``fluff'' to the southeast falls within the larger circle, but the smaller $r=0.25^{\prime\prime}$ aperture avoids excessive contamination.}
\label{fig:spec_model}
\end{figure*}

The main target of this paper, \galaxylong, was selected by searching for galaxies with strong Balmer breaks in public spectra on the Dawn JWST Archive\footnote{https://dawn-cph.github.io/dja/} \citep[DJA;][]{brammer_2023_8319596,Heintz2024, DeGraaff2025}, an online repository which contains reduced image mosaics, photometric catalogs, and reduced NIRSpec MSA spectra from public JWST programs. Unlike most other grating spectra for $z\gtrsim2$ quiescent galaxies which commonly have Balmer breaks, \galaxylong\ features a prominent $4000\ \mathrm{\AA}$ break with deep Ca H\&K absorption visible in the NIRSpec G140M spectrum and stellar metal absorption lines in the G235M spectrum (Figure \ref{fig:spec_model}). There is a distinct lack of nebular emission lines in the spectrum aside from faint \NII\ and \SII\ emission, which could be produced by a low-luminosity active galactic nucleus (AGN) or evolved, low-mass post-asymptotic giant branch stars \citep[e.g.,][]{Binette1994, Byler2019}. These features visually suggest a very old ($> 1$ Gyr) stellar population that first formed when the Universe was young. Based on its suspected early formation time and apparent quiescent nature, we nickname it ``Eridu,'' after the ancient Bronze Age Sumerian city in Mesopotamia.

\galaxy\ lies at R.A. $03^\mathrm{h}:32^\mathrm{m}:24.48^\mathrm{s}$ and Dec. $-27^\circ:50^\prime:04.65^{\prime\prime}$ in the GOODS-South field, one of the two fields targeted by the Great Observatories Origins Deep Survey \citep[GOODS][]{Giavalisco2004}. Further extensive HST imaging from the Cosmic Assembly Near-Infrared Deep Extragalactic Survey \citep[CANDELS;][]{Grogin2011,Koekemoer2011} and the Hubble Ultra Deep Field \citep[HUDF;][]{Beckwith2006,Bouwens2010,Ellis2013} has led to the status of GOODS-S as one of the most-studied extragalactic legacy fields.

The JWST photometry used in this work largely come from the JWST Advanced Deep Extragalactic Survey \citep[JADES, PIDs 1180, 1210, 1286, 3215;][]{Eisenstein2023a} and the Systemic Mid-infrared Instrument Legacy Extragalactic Survey \citep[SMILES, PID 1207;][]{Rieke2024}. GOODS-S also benefits from imaging and parallels from several other early JWST programs including the First Reionization Epoch Spectroscopic Complete program \citep[FRESCO, PID 1895;][]{Oesch2023}, the JWST Extragalactic Medium-band Survey \citep[JEMS, PID 1963;][]{Williams2023}, and the program `Quiescent or dusty? Unveiling the nature of red galaxies at $z>3$' \citep[PID 2198;][]{Barrufet2025}.

\subsection{HST and JWST Photometry} \label{ssec:img}

In this work, we use the public JADES GOODS-South v2.0 photometric catalog (hereafter, just the JADES catalog), downloaded from MAST\footnote{https://archive.stsci.edu/hlsp/jades, doi:10.17909/8tdj-8n28. Imaging from JEMS (doi:10.17909/fsc4-dt61) is included in the JADES GOODS-S catalog.}. This catalog was constructed using the general JADES imaging reduction and source extraction process outlined in \citet{Rieke2023, Hainline2024} with the new imaging data in JADES NIRCam second data release \citep{Eisenstein2023a}. 

There are 20(!) measured fluxes in each filter per source in the JADES catalog, with 10 aperture options and two aperture correction options. Fluxes are measured in seven circular fixed apertures (six at $r=0.10^{\prime\prime}$, $0.15^{\prime\prime}$, $0.25^{\prime\prime}$, $0.30^{\prime\prime}$, $0.35^{\prime\prime}$, and $0.50^{\prime\prime}$ and one aperture containing 80\% of the encircled energy as determined by the point spread function (PSF) in each filter), two Kron apertures (with $K$ parameters $K=2.5$ and $K=1.2$), and finally as all of the pixels in the segmentation map summed together. For each aperture, aperture correction is determined via two different methods: a) the aperture correction is calculated in a given filter using that filter's empirical PSF b) or the filter's mosaic is PSF-matched to the F444W filter's mosaic.

In this work, we use the PSF-matched fluxes from the $r=0.25^{\prime\prime}$ circular aperture (the white circle in the inset of Figure \ref{fig:spec_model}). This is approximately half of the Kron aperture's radius ($r=0.56^{\prime\prime}$) for this source. While Kron apertures are often the photometric standard for extended sources (e.g., galaxies), the presence of an additional source $\sim0.3$'' southeast of the target (see the inset in Figure~\ref{fig:spec_model}) motivates us to use a smaller aperture to minimize contamination by the neighboring object. This extra ``fluff'' could be a foreground source or an interacting object, but because it is not identified as an independent source in the JADES catalog and has no available spectroscopy, we exclude it from our analyses. The light within the $r=0.25^{\prime\prime}$ aperture also better matches the size and region probed by the spectrum (magenta boxes in Figure \ref{fig:spec_model} show MSA shutter locations).

We match our source's entry in the JADES catalog with the eight-band SMILES MIRI photometric catalog \citep[][downloaded from MAST\footnote{https://archive.stsci.edu/hlsp/smiles, doi:10.17909/et3f-zd57}]{Alberts2024}. The SMILES catalog was produced using the same pipeline as the JADES catalog and largely following almost identical methods, with custom background subtraction techniques and different aperture options setting the SMILES catalog apart. In F560W and F770W aperture corrections are calculated using empirical PSFs from high-SNR commissioning data, and at $\gtrsim10\mu m$ model PSFs are generated with \texttt{WebbPSF}. Fluxes are measured in circular $r=0.25^{\prime\prime}$, $0.3^{\prime\prime}$, $0.35^{\prime\prime}$, $0.5^{\prime\prime}$, and $0.6^{\prime\prime}$ apertures and a $K=2.5$ Kron aperture. We opt to use the smallest SMILES aperture ($r=0.25^{\prime\prime}$) to match our choice in the JADES catalog, but we note that the aperture-corrected fluxes in the SMILES catalog for \galaxy\ are all quite consistent with each other.

In total, \galaxy\ is observed in 27 photometric space-based filters from $0.4-25\ \mu\mathrm{m}$: 
\begin{itemize}
    \item five HST/ACS bands: F435W, F606W, F775W, F814W, and F850LP;
    \item four HST/WFC3 bands: F105W, F125W, F140W, and F160W;
    \item 10 JWST/NIRCam bands: F090W, F115W, F150W, F182M, F200W, F210M, F277W, F356W, F410M, and F444W;
    \item and eight JWST/MIRI bands: F560W, F770W, F1000W, F1280W, F1500W, F1800W, F2100W, and F2550W.
\end{itemize}We show the HST, JWST NIRCam, and JWST/MIRI photometry and the full spectral energy distribution (SED) in $f_\nu$ units for \galaxy\ in the top of Figure \ref{fig:phot_model} (HST and JWST/NIRCam photometry from $0.4-5~\mu$m is shown in $f_{\lambda}$ units at the bottom).

\subsection{JWST Spectroscopy} \label{ssec:spec}

\galaxylong\ was observed with JWST/NIRSpec MOS by the SMILES JWST program in 2023 August for 7000 s in the medium-resolution G140M and G235M grating dispersers with the F100LP and F170LP transmission filters, respectively. Target selection and a general overview of the NIRSpec observations is presented in \citet{Alberts2024} and a full description will be presented in Y. Zhu et al. (2025, in preparation). 

Reduced science spectra for \galaxy\ were downloaded from the DJA. In this work, we use the DJA v3 reduction of this galaxy's spectra, which covers the standard wavelength ranges for the G140M/F100LP (0.97–1.84 $\mu$m) and G235M/F170LP (1.66-3.07 $\mu$m) spectra (hereafter, just G140M and G235M spectra). Public spectra hosted on the DJA were reduced using the single standardized \texttt{msaexp} pipeline described in detail in \citet{Morishita2023a, DeGraaff2024a, Heintz2024}. The \texttt{msaexp} pipeline follows the standard JWST reduction with additional custom routines to remove bias on individual exposures, mask snowballs, and correct for $1/f$ noise in the spectra. The final 1D science spectrum and associated uncertainties are extracted from the rectified 2D image using the \citet{Horne1986a} optimal extraction procedure.

\subsubsection{Cosmic Ray Contamination} \label{sssec:CR}

A bright emission feature appears in the G140M spectrum of \galaxy\ at $17115-17200$ \AA\ (this feature also appears in the automatic MAST science product for this galaxy). This feature is close to the wavelength of \HeII, so the required blueshift and the slight spatial offset from the continuum could indicate a remarkable outflow from a powerful AGN. However, other emission lines often associated with AGN are either weak (\NII\ and \SII\ doublets) or completely undetected (\OIII\ doublet and H$\alpha$). Therefore, before we perform any further analysis of this galaxy, we investigate the origin of this feature.

\galaxy\ falls on the right detector (NRS2, Quadrant 2) and there are no additional open shutters along the direction of dispersion. Therefore, we can definitively rule out contamination from the zero-order image of another source to its right. Open shutters on NRS1 directly to the left of \galaxy\ on the detector fall on blank sky and are empty, so we conclude that \galaxy\ is also uncontaminated by first or higher order emission from a bright galaxy to its left. By matching the mask slits to the ``rate'' Level 2 files downloaded from MAST and visually examining the slit centered on \galaxy, we find a clear continuum signal and visible Ca H\&K absorption. In the first exposure (file ``jw01207004001\_03101\_00001\_nrs2\_rate.fits'') a cosmic ray incident on the continuum of \galaxy\ was incompletely masked and lies roughly at the location of the feature in question in the reduced 2D spectrum. We find no evidence of emission at this point in either of the other Level 2 images. Therefore, we conclude that this feature is contamination from a cosmic ray and we mask out the 17115-17200\AA\ region in all of our analyses.

\subsubsection{Combined spectrum} \label{sssec:combo}

We combine the G140M and G235M spectra using a method based on \citet{Carnall2024}. First, we visually inspect the two grating spectra in the overlap region for a consistent spectral shape. We find that they are consistent so we do not apply a polynomial re-calibration. We do, however, scale the average integrated fluxes from the spectra in the overlap region. The integrated flux of the G235M grating spectrum is 0.992 times that of the G140M in this region, so we apply this minor correction to the entire G235M spectrum. Using \texttt{SpectRes} \citep{Carnall2017}, we degrade the G140M spectrum to the resolution of the G235M spectrum in the overlap region. Finally, we add the two overlap spectra and their error spectra, weighted by their inverse variance. We use this final combined spectrum (black line in Figure \ref{fig:spec_model}) in all subsequent analyses shown in this work unless noted otherwise. 

\section{Prospector Modeling} \label{sec:modeling}

To infer the stellar population properties and recover the SFH of \galaxy, we simultaneously fit the HST/JWST photometry and NIRSpec spectrum of this galaxy with the Python-based Bayesian SED modeling tools \texttt{Prospector} v1.4 \citep{Johnson2017, Leja2017, Johnson2021} with the nested sampling code \texttt{dynesty} \citep{Speagle2020}.
We detail the models and prior choices below (all priors are uniform unless noted otherwise).

\subsection{The Model}\label{ssec:p_model}

At its core, \texttt{Prospector} fits observed photometric and spectroscopic data to a model galaxy constructed from the synthetic stellar population library FSPS \citep{Conroy2009a, Conroy2010a}. We use the MIST isochrones \citep{Choi2016, Dotter2016}, MILES spectral stellar library, and a Chabrier initial mass function \citep{Chabrier2003a}.

Our \texttt{Prospector} model is constructed similar to that of other works studying quiescent galaxies \citep[e.g.,][]{DeGraaff2024a, Weibel2025, Park2024a, Turner2025, Slob2024}. We allow the redshift to vary $z_{\mathrm{spec}}\in[z_\mathrm{spec}-0.05, z_\mathrm{spec}+0.05]$, where $z_\mathrm{spec}$ is the best-fit redshift in the DJA spectroscopic catalog. The total mass formed is set $\log M_{*,\mathrm{formed}}/\mathrm{M_\odot}\in[7,12].$
We fit stellar metallicity with a uniformly sampled logarithmic prior $\log Z/\mathrm{Z_\odot}\in [-1, 0.19]$. We use a two-parameter \citet{Kriek2013} dust law with a free optical depth $\tau\in[0, 4]$ and deviation from the \citet{Calzetti2000a} dust law slope $\delta\in[-1, 0.4]$. Attenuation around young stars ($t<10$ Myr) is fixed to be twice that of older populations. Because the SMILES photometry covers the 6.2 and 7.7 $\mu$m PAH features (Figure \ref{fig:phot_model}), we also include the \citet{Draine2007} dust emission model with parameters $\gamma_{\mathrm{e}}\in[0, 0.15]$, $U_{\mathrm{min}} \in[0.1, 15]$, $q_{\mathrm{PAH}}\in[0.1, 10]$ as nuisance parameters. 

 \texttt{Prospector} models nebular emission lines in one of two ways: it can use the CLOUDY nebular emission grids in FSPS computed for stellar ionization sources (i.e., emission lines powered by young stars), or it can independently fit emission lines in the residual spectrum with no assumptions as to the underlying cause. As faint \NII\ and \SII\ emission is visible in the G235M spectrum, we apply the latter option with the nebular marginalization procedure in our fiducial model to avoid ascribing  emission from potential AGN activity or post-asymptotic giant branch stars to star formation. Choosing the first option and allowing physical nebular emission to infill Balmer absorption lines does not change our conclusions, though the best-fit \texttt{Prospector} models are unable to recreate the faint \NII\ and \SII\ emission.

 To account for the medium resolution of the gratings, we convolve our models with the JDOX resolution curve multiplied by a factor of 1.3 \citep{Curtis-Lake2023, DeGraaff2024a, Nanayakkara2024, Slob2024}. With the prominent absorption in \galaxy's spectrum, we fit for the continuum velocity dispersion with a wide prior $\sigma_{*}\in[0,1000]~\rm{km\ s^{-1}}$. We also convolve the fit emission lines to fit the gas velocity dispersion $\sigma_{\rm{gas}}\in[0,1000]~\rm{km\ s^{-1}}$ (assuming all emission lines have the same width).

To flux-calibrate the spectrum, we utilize the \texttt{PolySpecModel} procedure. With this option, \texttt{Prospector} accounts for deviations between the observed spectrum and photometry by multiplying an $n$th degree Chebyshev polynomial to the spectrum to match it to the model at each likelihood call. We use $n=10$, though lower-degree polynomials ($n=2$) produce similar results, albeit with higher $\chi^2$ values and wavelength-dependent residuals.

We include an outlier model and noise jitter term to mitigate mismatch between the model and observed data, bad pixels, and underestimated noise. The outlier model fits a fraction of spectroscopic outlier data points $f_{\rm{out}}\in[10^{-5},0.2]$ and inflates their uncertainties by a factor of 5. The spectroscopic noise jitter term $j_{\rm{spec}}\in[0.5,3.0]$ multiplied to all uncertainties is fit by including kernels for uncorrelated noise in the likelihood calculations \citep{Maseda2023}.

Finally, we utilize a nonparametric continuity SFH \citep{Leja2019} with 14 SFH bins. We include three young bins at [0, 10 Myr], [10 Myr, 50 Myr], and [50, 100 Myr] to recover any recent change in star formation of \galaxy. In preliminary testing of our model, after the three young bins we tried both logarithmically spaced bins and uniformly spaced bins until $t_\mathrm{obs}$ (where $t_\mathrm{obs}=2.47\ \mathrm{Gyr}$, the age of the Universe at the observed redshift of \galaxy), but the inferred SFHs with this scheme placed all of the star formation in the earliest bin, necessitating finer subdivision of SFH bins at high redshift. Therefore, we split the remaining 11 bins into 5 logarithmically spaced bins until $0.6t_\mathrm{obs}$ and then two equally spaced bins $0.1t_\mathrm{obs}$ wide and four equally spaced bins each $0.05t_\mathrm{obs}$ wide (approximately $250$ and $125$ Myr, respectively). We apply a Student's t-distribution prior centered at $\Delta \log \mathrm{SFR} = 0$ with $\sigma=0.3$ and $\nu=2.0$ between neighboring bins as in \citet{Leja2019}.

\subsection{Analysis} \label{sec:p_analysis}

\begin{figure*}[!htb]
\centering
\includegraphics[width=\linewidth]{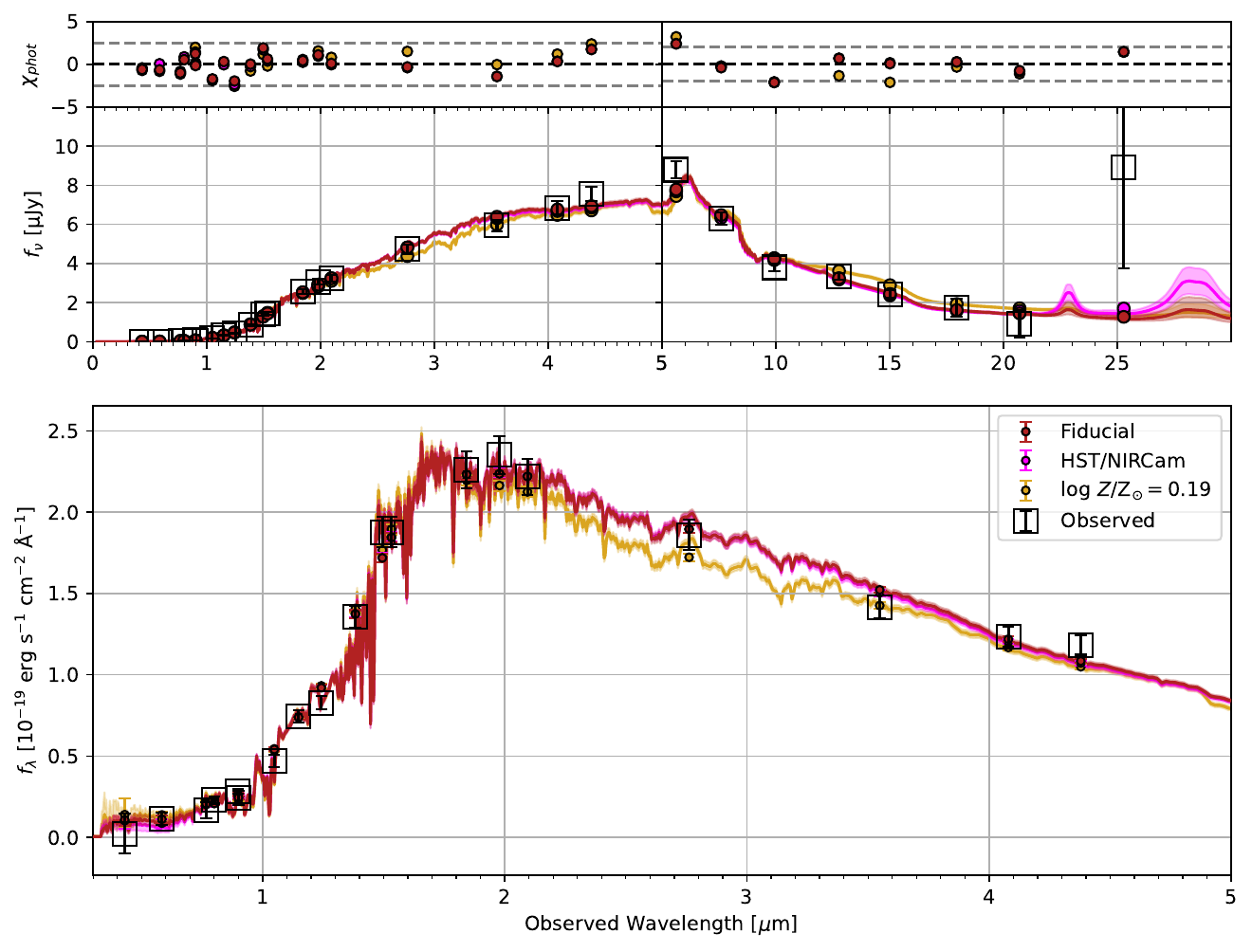}
\caption{We demonstrate how our Prospector models match the observed photometry. We show the fiducial model including (red) and excluding (magenta) MIRI photometry and the model using with metallicity fixed to $\log Z/\mathrm{Z_\odot}=0.19$ (goldenrod). The shaded region shows the $1\sigma$ uncertainties in the model posteriors. excluding the MIRI data does not noticeably effect the inferred SED even in the $5-25\ \mu$m range. All models reliably reproduce the break, but redward of $4000\ \mathrm{\AA}$ the fixed metallicity model visibly diverges from the first two. The later-forming fixed metallicity model has larger residuals than the faster-forming models. This is particularly noticeable in the rest-frame near-infrared (i.e., F277W), though the SEDs of all models exhibit some difficulty reproducing the slight observed bump at $\sim2\ \mu$m and red $3-5\ \mu$m colors even with a flux uncertainty floor of $5\%$ applied to the photometry.
}
\label{fig:phot_model}
\end{figure*}

For our \texttt{Prospector} modeling, we fit the $0.4-25~\mu$m photometry from the JADES and SMILES catalogs in the $0.25^{\prime\prime}$ aperture and the combined NIRSpec G140M+G235M spectrum. We apply a S/N ceiling of 20 to the uncertainty of the observed photometry to account for systematic uncertainties in the stellar libraries. In the spectrum, we mask Na ID absorption in G235M, the cosmic ray in G140M, and a short isolated section of the G140M with $\lambda_{\rm obs}<10000$\AA. Because this last section of spectrum has low S/N, is discontinuous from the rest of the spectrum, and lies in the low resolution wavelength range of MILES stellar library, removing it does not affect our results. To account for difficulty in sampling the underlying multi-modal posterior, we fit each model multiple times and use the highest average likelihood run in our analysis (see Appendix~\ref{sec:sampling} for details).

Our fiducial \texttt{Prospector} modeling confirms that \galaxy\ is massive and quiescent. We infer a stellar mass $\log(M_*/\mathrm{M_\odot}=10.96^{+0.01}_{-0.01})$, a $3\sigma$ upper limit on average star formation rate (SFR) over the last 100 Myr of $\mathrm{SFR_{100}}<1.54\ \mathrm{M_\odot\ yr^{-1}}$ , and a $3\sigma$ upper limit on specific star formation rate $\log(\mathrm{sSFR_{100}\ yr})<-10.77$. Other inferred properties of \galaxy\ are presented in Table~\ref{tab:megatable1}. The best-fit (MAP) \texttt{Prospector} model fit to the combined spectrum and photometry is shown as a red line in Figure \ref{fig:spec_model}, with the observed spectrum scaled to the model and photometry by the calibration polynomial described in \ref{ssec:p_model}. We show the model from $0.3-30~\mu$m with the observed photometry in Figure \ref{fig:phot_model}. The model does a good job reproducing the observed photometry, but has some difficulty matching the slight bump at $2\ \mu$m and red $3-5\ \mu$m colors.

\begin{figure*}[!htb]
\centering
\includegraphics[width=\linewidth]{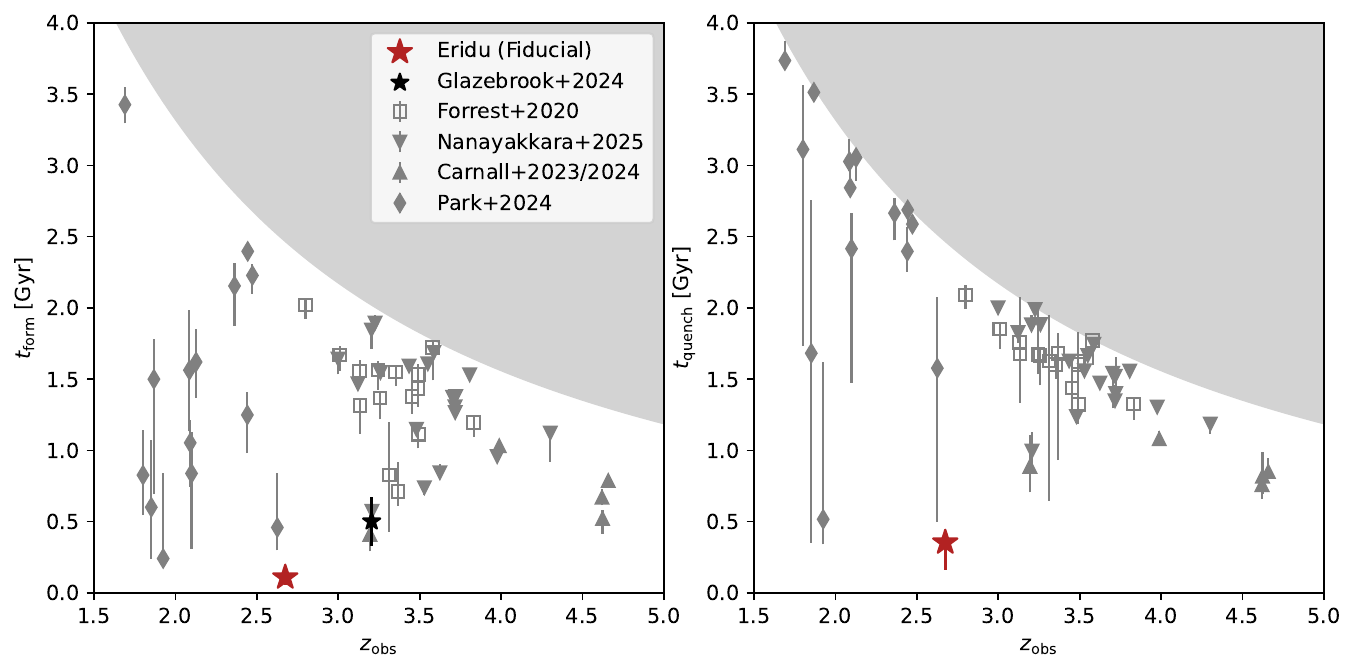}
\caption{In the left panel we show $t_{\mathrm{form}}$ and in the right $t_{\mathrm{quench}}$ as a function of $z_{\mathrm{spec}}$ for \galaxy\ (red star) and other massive quiescent galaxies from the literature \citep{Forrest2020, Nanayakkara2024, Nanayakkara2025, Carnall2023a, Carnall2024, Park2024, Glazebrook2024}. The red star indicates the $t_{\mathrm{form}}$ and  $t_{\mathrm{quench}}$ of our fiducial model, which infers earlier formation and quenching than any other source.
}
\label{fig:times}
\end{figure*}

\begin{figure*}[!htb]
\centering
\includegraphics[width=\linewidth]{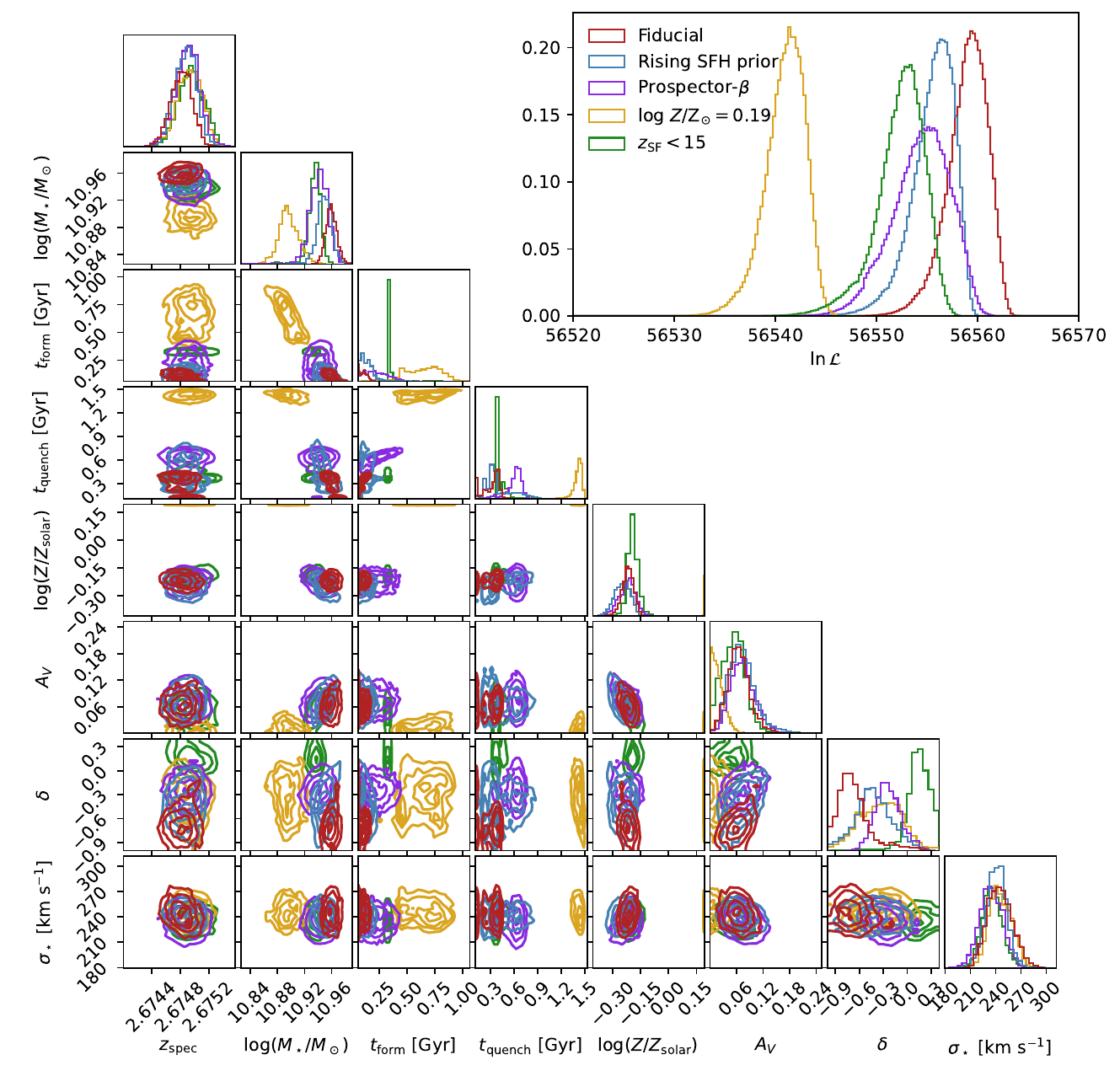}
\caption{Inferred posteriors for select \texttt{Prospector} quantities and for a subset of the models fit to the data. We show excellent agreement between models for redshift, stellar mass, formation and quenching times, metallicity, dust attenuation, deviation from the \citet{Calzetti2000a} dust law slope, and stellar velocity dispersion (more quantities for each model are given in Tables~\ref{tab:megatable1} and \ref{tab:megatable2}). The main outlier model shown here (fixed $\log Z/\mathrm{Z_\odot}=0.19$) also corresponds with a lower $\ln\ \mathcal{L}$ sampled distribution (weighted histogram of log likelihoods is shown in the top left). Notably, posteriors with later inferred $t_\mathrm{form}$ tend to have lower $\ln\ \mathcal{L}$ too (see Table~\ref{tab:megatable2} as well).}
\label{fig:posterior}
\end{figure*}

To quantify the star formation history of \galaxy, we define two parameters, $t_{\rm{form}}$ and $t_{\rm{quench}}$, as the age of the universe at which the galaxy formed 50\% and 90\% of its stellar mass, respectively (shown in Table~\ref{tab:megatable2} and as a red star in Figure~\ref{fig:times}). Our \texttt{Prospector} fit points to an extremely early formation time and quenching, with 
with $t_{\rm{form}}=108_{-39}^{+37}$ Myr and $t_{\rm{quench}}=353_{-179}^{+53}$ Myr ($z_{\rm{form}}=28.41_{-5.29}^{+10.31}$	and	$z_{\rm{quench}}=12.46_{-1.20}^{+8.10}$). We compare \galaxy's $t_{\rm{form}}$ and $t_{\rm{quench}}$ with other massive quiescent galaxies from the literature in Figure~\ref{fig:times}. The $t_{\rm{form}}$ inferred with our fiducial \texttt{Prospector} model (the red star) is notably earlier than other old, quiescent sources in the literature, even the ancient $z=3.2$ galaxy ZF-UDS-7329 \citep[][the black star;]{Glazebrook2024}. A corner plot with $\log M_*/\mathrm{M_{\odot}}$, $t_{\rm{form}}$, $t_{\rm{quench}}$, $\log Z/\mathrm{Z_\odot}$, dust extinction $A_V$, and other properties inferred with this model is shown in Figure \ref{fig:posterior} in red.

\begin{figure*}[!htb]
\centering
\includegraphics[width=\linewidth]{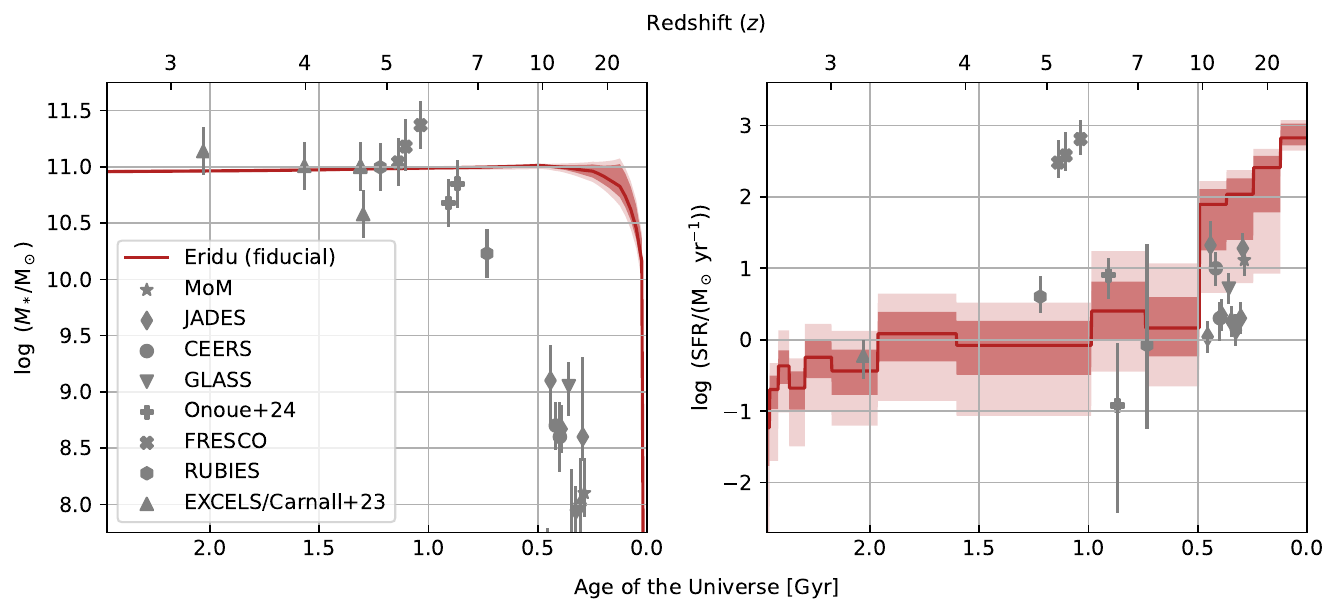}
\caption{The SFH of \galaxy\ inferred from our fiducial \texttt{Prospector} model. On the left we show stellar mass and on the right SFR as a function of cosmic time/redshift. The fiducial model's SFH is shown with the median, $1\sigma$ uncertainty, and $2\sigma$ uncertainty represented by the solid colored line, dark shaded, and light shaded regions, respectively. For comparison we include stellar mass and SFR measurements from the literature for spectroscopically confirmed sources at $z>10$: MoM-14 \citep{Naidu2025}, JADES-GS-z14-0 and -z14-1 \citep{Carniani2024}, JADES-GS-z10, -z11, -z12, and -z13 \citep{Curtis-Lake2023}, JADES-GN-z11 \citep{Tacchella2023}, Maisie's galaxy and CEERS2\_588 \citep{ArrabalHaro2023}, and GHZ2/GLASS-z12 \citep{Castellano2024}. We also show several prominent massive and/or quiescent galaxies at high redshift: two post-starburst quasar hosts \citep{Onoue2024}, three massive star forming galaxies from FRESCO \citep{Xiao2024}, GS-9209 \citep{Carnall2023a}, RUBIES-UDS-QG-z7 \citep{Weibel2025}, RUBIES-EGS-QG-1 \citep{DeGraaff2024a}, and the four ultra-massive quiescent galaxies from EXCELS \citep[][two which were first presented in \citealt{Glazebrook2024} and \citealt{Nanayakkara2024}]{Carnall2024}. We apply a minimum uncertainty of 0.2 dex to the SFRs and stellar masses of sources from the literature to account for different assumptions in SED modeling \citep{Conroy2013}. While the fiducial model's inferred SFH is unremarkable when compared with the stellar mass and SFR of other $z<10$ galaxies, at $z>10$ the model predicts that \galaxy's progenitor is nearly a factor of $\sim100$ times more massive than some of the brightest and most star forming galaxies observed at cosmic dawn.}
\label{fig:sfh_fiducial}
\end{figure*}

In Figure \ref{fig:sfh_fiducial}, we show the star formation history (SFH) of \galaxy\ inferred by the fiducial model. The left panel shows stellar mass and the right panel shows star formation rate, both as functions of cosmic time/redshift. We include spectroscopically confirmed high redshift massive galaxies \citep{Carnall2024, DeGraaff2024a, Xiao2024, Onoue2024, Weibel2025} and $z>10$ galaxies \citep{Castellano2024, Tacchella2023, ArrabalHaro2023, Curtis-Lake2023, Carniani2024, Naidu2025} from the literature for comparison. While the inferred star formation rates at $z<15$ generally agree with observations, all galaxies observed at cosmic dawn to date are only beginning to assemble their stellar mass (i.e., their star formation rates are increasing), meanwhile the bulk of \galaxy's stars have already been formed and star formation is declining. The extremely fast and early formation inferred with this model produces a $\log M_*/\mathrm{M_\odot}>11$ galaxy by $z\sim20$, more massive than most $z>10$ observations by over two orders of magnitude! Therefore, we explore whether altering our \texttt{Prospector} model can better match high-redshift observations.

\subsection{Alternative Fits}
\label{ssec:altmodels}

To further investigate the formation history of \galaxy, we test how robust the recovered formation time is to our choice of model and observed data by altering these inputs. We show our results for all these models in Tables~\ref{tab:megatable1} and \ref{tab:megatable2}.

\subsubsection{Altering Input Observations}

As a test of whether the early formation time of our fiducial model is driven by any particular feature in the observed data, we refit \galaxy\ with the same model and using different subsets of the photometry and spectroscopy.

\textbf{Without MIRI:} First, we run our fiducial model (described in \S\ref{ssec:p_model}) on the input data with MIRI photometry masked out (i.e., utilizing only fluxes from the JADES HST and JWST/NIRCam catalog). We still use the dust emission template even though all three parameters are completely unconstrained in this run. The SED fit from this run is shown in magenta in Figure \ref{fig:phot_model}.

\textbf{G140M or G235M only:} Next, we fit the fiducial model but swap out the combined G140M+G235M spectrum for the individual G140M or G235M spectra to test whether the early formation time is driven by any one particular feature.

We find that changing the observed data modeled with \texttt{Prospector} in this way produces results consistent with the initial run with the complete input dataset: extremely and rapid formation. 

\subsubsection{Disallowing High-Redshift Star Formation}

Because the fiducial model's inferred SFH would imply that \galaxy's progenitor is over 100 times more massive than the already rare, bright, and highly star-forming $z>10$ galaxies seen in observations, we modify the SFH binning to investigate whether forbidding earlier star formation produces inferred formation histories that better match high redshift observations.

$\mathbf{z_{\mathrm{\bf{SF}}}}$: In this model variant, we alter our fiducial model's star formation history bin edges. As in \citet{Turner2025}, we remodel \galaxy\ disallowing any star formation in three different epochs: $z>20$, $z>15$, and $z>10$. For these fits, we rescale the SFH bin spacing according to the onstart of star formation (i.e., bins are spaced according to $t(z_{\rm{SF}})$ rather than $t_\mathrm{obs}$).

As with the fiducial model, the bulk of star formation occurs in the earliest time bin, though moving the bins causes $t_{\rm{form}}$ and $t_{\rm{quench}}$ to shift accordingly. Though the star formation cutoffs span a large redshift range, the difference in cosmic time between $z=10$ and the Big Bang is only $\sim500$ Myr, so the relative differences between the various $z_{\rm{SF}}$ models and our fiducial model are minor.
We show the $z_{\rm{SF}} < 15$ model as green in Figures \ref{fig:posterior} and \ref{fig:sfh_alternative}.

\subsubsection{Different SFH Priors}

The default Student's t-distribution prior on the continuity SFH from \citet{Leja2019} encourages SFR ratios between bins to trend toward unity (i.e., the model prefers SFR remain constant between the onset of star formation and the time of observation). However, this choice of prior a) has difficulty reproducing the bursty star formation histories expected in the high-redshift universe and b) makes no distinction for cosmic trends of stellar mass assembly. Therefore, we test alternative priors on our fiducial model. 

\textbf{Bursty Prior:} First, we implement the \citet{Tacchella2022} ``bursty'' variant of the continuity prior. This prior widens the Student's t-distribution by increasing $\sigma$ from 0.3 to 1.0, which effectively allows for more extreme variation between star formation rates in neighboring bins.

As might be expected based on the previous modeling, \texttt{Prospector} once again puts the majority of the model's star formation in the earliest time bin. By widening the Student's t-distribution and allowing SFH bins to undergo greater changes in SFR, the model is free to prefer this extremely early formation mode.

\textbf{Rising SFH prior:} To account for the expectation that SFRs should rise as galaxy halos assemble more mass and allow for the conversion of more baryons into stars, \citet{Turner2025} scaled the mean star formation each continuity bin to the average accretion rate of dark matter halos in an Einstein-deSitter cosmology \citep{Dekel2013}, which is a good approximation for the high-redshift universe. We apply the scaling derived therein as a prior on our SFH. This model is shown in blue in Figures \ref{fig:posterior} and \ref{fig:sfh_alternative}.

\textbf{Prospector-$\beta$:} We also implement the Prospector-$\beta$ SFH prior from \citet{Wang2023}. Similar to the \citet{Turner2025} rising SFH prior, the Prospector-$\beta$ prior modifies the expectation value in each star formation bin to match the cosmic star formation rate densities from \citet{Behroozi2019}. The priors are further modified by a mass dependence, such that more massive galaxies form earlier. Stellar mass is subject to a prior based on the galaxy stellar mass function and galaxy number density. Finally, the stellar mass-stellar metallicity relationship is included in this prior as well. We direct the reader to \citet{Wang2023} for technical details of the underlying assumptions. For our purposes here, this means that the Prospector-$\beta$ SFH prior is a function of stellar mass, which is in turn subject to a prior based on redshift, number density, and stellar metallicity. Therefore, this prior is more computationally expensive than a flat prior or prior which is simply a function of redshift \citep[e.g., the rising SFH prior from][]{Turner2025}. To reduce excessive posterior sampling runtimes, we shrink the size of our model's parameter space by adjusting the ranges of the formed stellar mass prior based on the posteriors from previous models: $\log M_\mathrm{*,\ formed}/\mathrm{M_\odot}\in [10.5, 11.5]$. This model is shown in purple in Figures \ref{fig:posterior} and \ref{fig:sfh_alternative}.

We find that the formation timescales inferred by the \citet{Turner2025} halo assembly and \citet{Wang2023} Prospector-$\beta$ priors are consistent with the fiducial model, with Prospector-$\beta$ preferring slightly later (but still remarkably early) formation and quenching.

\subsubsection{Other models}

\textbf{Kroupa IMF:} Though the shape of the Kroupa IMF \citep{Kroupa2001} is very similar to the Chabrier IMF \citep{Chabrier2003a}, we test how robust our modeling is to this choice. Results for the Kroupa model are indistinguishable from the fiducial Chabrier model. Based on the good agreement between IMFs, we assume the fiducial (Chabrier) model when drawing comparisons between our \texttt{Prospector} modeling and modeling with \texttt{Bagpipes} in \S\ref{sec:pipes} with \texttt{Bagpipes} and \texttt{alf$\alpha$} in \S\ref{sec:alfalpha} (both of which assume the Kroupa IMF).

$\mathrm{\textbf{log}}\ \mathbf{Z/}\mathrm{\textbf{Z}}\mathbf{_\odot=0.19}$: A high fixed metallicity model has been previously shown to infer later formation times \citep[e.g., ][]{DeGraaff2024a,Weibel2025}. Here, rather than leave metallicity as a free parameter we fix it to 2.5 times solar ($\log Z/\mathrm{Z_\odot}=0.19$). This model is shown in gold in Figures \ref{fig:phot_model}, \ref{fig:posterior}, \ref{fig:sfh_alternative}, and \ref{fig:progenitor_spec}.

\textbf{Single Burst}: Finally, we fit \galaxy\ with a simple stellar population (SSP), equivalent to all of the stars assembling in a single burst of star formation. The results of this test are discussed in Appendix~\ref{sec:ssp}.

\begin{deluxetable*}{cccccccc}
\centerwidetable
\tablewidth{0pt}
\tablecaption{Properties of \galaxy\ Inferred by SED Modeling \label{tab:megatable1}}
\tablehead{\colhead{Model} & \colhead{$z_\mathrm{spec}$} & \colhead{$\log M_*/\mathrm{M_\odot}$} & \colhead{$\mathrm{SFR_{100\ Myr}}$} & \colhead{$\log Z/\mathrm{Z_\odot}$} & \colhead{$A_V$} & \colhead{$\delta$} & \colhead{$\sigma_*$}\\
\colhead{} & \colhead{} & \colhead{} & \colhead{$[\mathrm{M_\odot\ yr^{-1}}]$} & \colhead{} & \colhead{} & \colhead{} & \colhead{$\mathrm{[km\ s^{-1}]}$}}
\startdata
Fiducial	&	$2.6748_{-0.0002}^{+0.0001}$	&	$10.96_{-0.01}^{+0.01}$	&	$0.31_{-0.13}^{+0.18}$	&	$-0.22_{-0.03}^{+0.03}$	&	$0.06_{-0.02}^{+0.03}$	&	$-0.70_{-0.14}^{+0.20}$	&	$244_{-13}^{+15}$\\
\hline
Without MIRI	&	$2.6748_{-0.0002}^{+0.0002}$	&	$10.96_{-0.01}^{+0.01}$	&	$0.17_{-0.10}^{+0.13}$	&	$-0.30_{-0.03}^{+0.03}$	&	$0.09_{-0.02}^{+0.02}$	&	$-0.38_{-0.13}^{+0.16}$	&	$240_{-13}^{+14}$\\
G140M	&	$2.6746_{-0.0002}^{+0.0002}$	&	$10.94_{-0.01}^{+0.01}$	&	$0.13_{-0.09}^{+0.11}$	&	$-0.18_{-0.06}^{+0.05}$	&	$0.06_{-0.03}^{+0.03}$	&	$-0.48_{-0.23}^{+0.22}$	&	$240_{-15}^{+15}$\\
G235M	&	$2.6748_{-0.0003}^{+0.0003}$	&	$10.95_{-0.01}^{+0.02}$	&	$0.13_{-0.10}^{+0.10}$	&	$-0.30_{-0.06}^{+0.04}$	&	$0.14_{-0.03}^{+0.04}$	&	$-0.31_{-0.42}^{+0.16}$	&	$255_{-39}^{+23}$\\
\hline
$z_\mathrm{SF}<10$	&	$2.6749_{-0.0002}^{+0.0002}$	&	$10.93_{-0.01}^{+0.01}$	&	$0.08_{-0.06}^{+0.07}$	&	$-0.18_{-0.03}^{+0.03}$	&	$0.06_{-0.03}^{+0.03}$	&	$-0.06_{-0.21}^{+0.17}$	&	$242_{-13}^{+13}$\\
$z_\mathrm{SF}<15$	&	$2.6750_{-0.0002}^{+0.0002}$	&	$10.94_{-0.01}^{+0.01}$	&	$0.06_{-0.05}^{+0.08}$	&	$-0.19_{-0.02}^{+0.03}$	&	$0.05_{-0.03}^{+0.03}$	&	$0.15_{-0.12}^{+0.13}$	&	$235_{-11}^{+12}$\\
$z_\mathrm{SF}<20$	&	$2.6749_{-0.0002}^{+0.0002}$	&	$10.95_{-0.01}^{+0.01}$	&	$0.10_{-0.07}^{+0.11}$	&	$-0.24_{-0.05}^{+0.05}$	&	$0.10_{-0.03}^{+0.03}$	&	$-0.14_{-0.23}^{+0.26}$	&	$234_{-12}^{+14}$\\
\hline
Bursty SFH	&	$2.6749_{-0.0002}^{+0.0002}$	&	$10.96_{-0.01}^{+0.01}$	&	$0.10_{-0.10}^{+0.26}$	&	$-0.24_{-0.04}^{+0.03}$	&	$0.06_{-0.03}^{+0.03}$	&	$-0.23_{-0.45}^{+0.34}$	&	$238_{-13}^{+12}$\\
Rising SFH	&	$2.6749_{-0.0002}^{+0.0001}$	&	$10.95_{-0.01}^{+0.01}$	&	$0.02_{-0.01}^{+0.01}$	&	$-0.25_{-0.05}^{+0.04}$	&	$0.07_{-0.03}^{+0.03}$	&	$-0.43_{-0.21}^{+0.24}$	&	$241_{-12}^{+11}$\\
Prospector-$\beta$	&	$2.6749_{-0.0002}^{+0.0002}$	&	$10.94_{-0.01}^{+0.01}$	&	$0.07_{-0.06}^{+0.11}$	&	$-0.22_{-0.05}^{+0.04}$	&	$0.07_{-0.02}^{+0.03}$	&	$-0.25_{-0.15}^{+0.16}$	&	$234_{-13}^{+13}$\\
\hline
Kroupa IMF	&	$2.6750_{-0.0001}^{+0.0002}$	&	$11.00_{-0.01}^{+0.01}$	&	$0.08_{-0.06}^{+0.07}$	&	$-0.23_{-0.06}^{+0.04}$	&	$0.07_{-0.03}^{+0.05}$	&	$-0.45_{-0.32}^{+0.43}$	&	$230_{-10}^{+12}$\\
$\log(Z/\mathrm{Z_\odot})=0.19$	&	$2.6749_{-0.0002}^{+0.0002}$	&	$10.90_{-0.01}^{+0.02}$	&	$0.16_{-0.09}^{+0.13}$	&	$0.19$	&	$0.02_{-0.01}^{+0.02}$	&	$-0.32_{-0.29}^{+0.25}$	&	$244_{-12}^{+14}$\\
\enddata
\tablecomments{Results from our \texttt{Prospector} fits to \galaxy\ for all of our models. For each model, we show the stellar mass ($\log(M_*/\mathrm{M_\odot})$), SFR averaged over the last 100 Myr ($\mathrm{SFR_\mathrm{100\ Myr}}$), metallicity ($\log(Z/\mathrm{Z_\odot})$), dust extinction ($A_V$), deviation from the \citet{Calzetti2000a} dust law slope ($\delta$), and stellar velocity dispersion $\sigma_*$.}
\end{deluxetable*}

\subsection{Results}

\begin{figure*}[!htb]
\centering
\includegraphics[width=\linewidth]{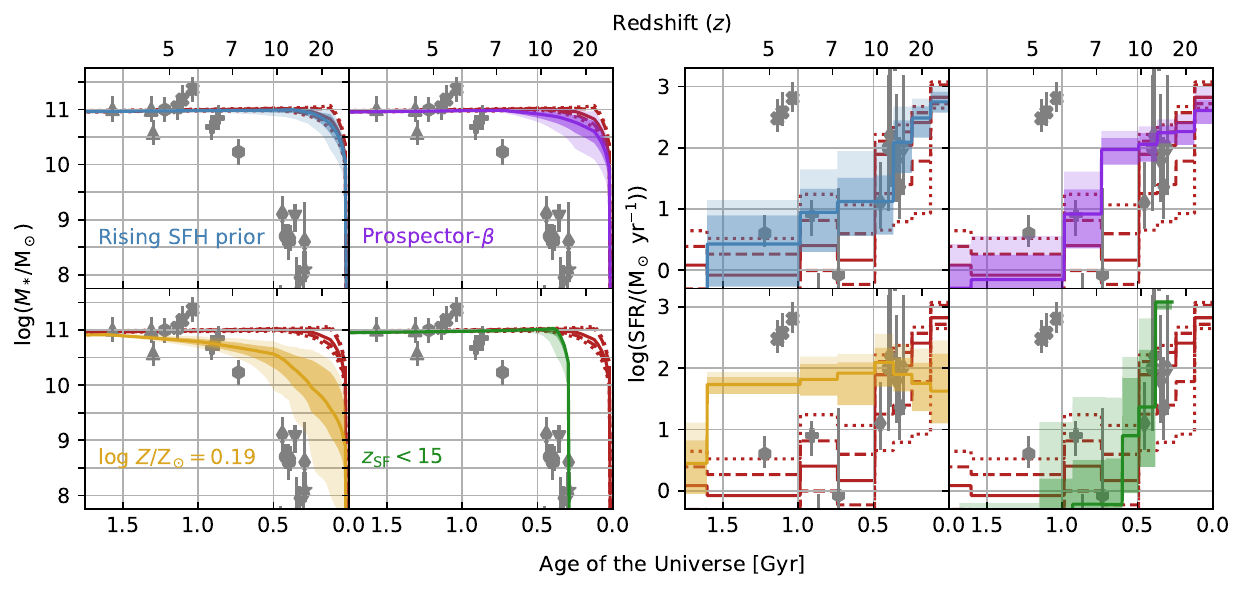}
\caption{The SFH of \galaxy\ inferred from a subset of our alternative \texttt{Prospector} models: the rising SFH \citep{Turner2025}, Prospector-$\beta$ \citep{Wang2023}, fixed metallicity, and $z_\mathrm{SF}<15$ (color-coded as Figure \ref{fig:posterior}). As with Figure~\ref{fig:sfh_fiducial} we show stellar mass on the left and SFR on the right, both as functions of cosmic time/redshift (here we have zoomed in on the time and SFR axes to focus on high redshift). In each of the alternative model panels we show the fiducial model's median/$1\sigma$/$2\sigma$ as solid/dashed/dotted red lines. As with the fiducial model, at $z>10$ the model SFH that \galaxy's progenitor is (at least) an order of magnitude more massive than brightest and most star forming galaxies observed at cosmic dawn. Fixing the model's metallicity to $\log Z/\mathrm{Z_\odot}=0.19$ (the maximum value we consider) reduces the burstiness of the inferred SFH and produces later $t_\mathrm{form}$, but still predicts masses and star formation rates in excess of high-redshift observations.}
\label{fig:sfh_alternative}
\end{figure*}

The alternative models infer the same trends as our fiducial model: \galaxy\ is a massive quiescent galaxy with sub-solar metallicity and is unobscured by dust. Inferred properties shown for all models in Table~\ref{tab:megatable1} and a subset of models are shown in Figure~\ref{fig:posterior}. As with the fiducial model, almost all of our alternative \texttt{Prospector} models also infer extremely early formation times (Table~\ref{tab:megatable2}). Only the fixed metallicity and $z_\mathrm{sf}<10$ models infer $t_\mathrm{form}>500\ \mathrm{Myr}$, and even then Figure~\ref{fig:sfh_alternative} demonstrates that the inferred SFH produces stellar masses at high redshift are still higher than the brightest spectroscopically-confirmed high-redshift galaxies (the stellar masses of which may be \emph{overestimated} if a top-heavy IMF is required to explain their extreme UV luminosities, e.g. \citealt{Hutter2025, Jeong2025}).

We also see that while the models which form later appear to better match (some) properties of high-redshift JWST observations, they fit the observed data for \galaxy\ worse with higher $\chi_\mathrm{phot}^2$ values (Table~\ref{tab:megatable2}). For each model, the highest likelihood MAP models tend to prefer faster formation timescales too, corresponding to higher initial star formation rates and earlier $t_\mathrm{form}$ and $t_\mathrm{quench}$ values. A trend of lower $t_\mathrm{form}$ and higher $\ln\ \mathcal{L}_\mathrm{avg}$ is also observed in comparisons between models overall. Repeat modeling of \galaxy\ (Appendix~\ref{sec:sampling}) and SSP tests (Appendix~\ref{sec:ssp}) further suggest that \texttt{Prospector} prefers a maximally old stellar population for \galaxy.

All of our modeling results in star formation rates in excess of $500\ \rm{M_{\odot}\ yr^{-1}}$ at extremely early cosmic times and in stellar mass in excess of $10^{11}\ \mathrm{M_\odot}$ at $z>10$ for all models (when allowed), except for fixed metallicity. The lower $2\sigma$ estimate for this slowest-forming model infers a mere $\log M_*/\mathrm{M_\odot}\approx10$ at $z=10$, which is still a poor match to observations. If the inferred SFHs are accurate, one would expect that such a progenitor would clearly be detectable by JWST photometry and spectroscopy.

\subsubsection{Do Such Progenitors Exist?}

Motivated by the extremely high stellar mass and star formation rates at early cosmic times inferred by our \texttt{Prospector} modeling, we set out to answer the questions, ``how would the spectrum of said progenitor compare to high-redshift observations?'' and ``would the progenitor of such a galaxy be easily detectable in JWST photometry?'' To answer these questions, we alter our \texttt{Prospector} model to predict a progenitor galaxy.

We modify our baseline fiducial model to define a ``progenitor model.'' First, to produce nebular emission lines from star formation at high redshift, we turn off the nebular marginalization procedure (i.e., the model produces emission lines based on the CLOUDY grids in FSPS). We also account for absorption due to the intergalactic medium using the model from \citet{Madau1995} (the updated \citealt{Inoue2014} model has not been implemented in FSPS, but the difference in IGM attenuation between models is not significant enough to change our conclusions). Because our progenitor model is not fit to observed data, we remove the spectroscopic jitter, spectroscopic outlier, and polynomial calibration, and we do not convolve the model's emission or absorption features with any grating resolution curve. Finally, we treat the star formation bins defined in \S\ref{ssec:p_model} as fixed in cosmic time and truncate them accordingly depending on the redshift at which the model is evaluated (for example, at $z=20$, the progenitor model contains all of the first earliest bin and less than half of the second).

We derive a distribution of quantities $\mathrm{SFR}(z)$, and $M_{*,\mathrm{formed}}(z)$ from the inferred posterior. \texttt{Prospector} does not model the metallicity evolution of the stellar population,
so we set the progenitor's metallicity to scale linearly with the mass formed (i.e., $Z(z)/Z(z_{\mathrm{obs}}) = M_{*,\mathrm{formed}}(z)/M_{*,\mathrm{formed}}(z_{\mathrm{obs}})$), with a minimum $\log Z/\mathrm{Z_\odot}=-1$. We do not introduce a time-dependence to the dust model parameters\footnote{This strong assumption produces a progenitor with minimal dust obscuration. If dust extinction set to $A_V=2,3,4$, the rest-frame UV is sufficiently faint that a progenitor would be nearly undetectable at $z>10$ in all NIRCam bands except for F444W.} or stellar/gas velocity dispersion (i.e., these parameters are fit to the inferred $z_{\mathrm{obs}}$ parameters).

\begin{figure*}[!htb]
\centering
\includegraphics[width=\linewidth]{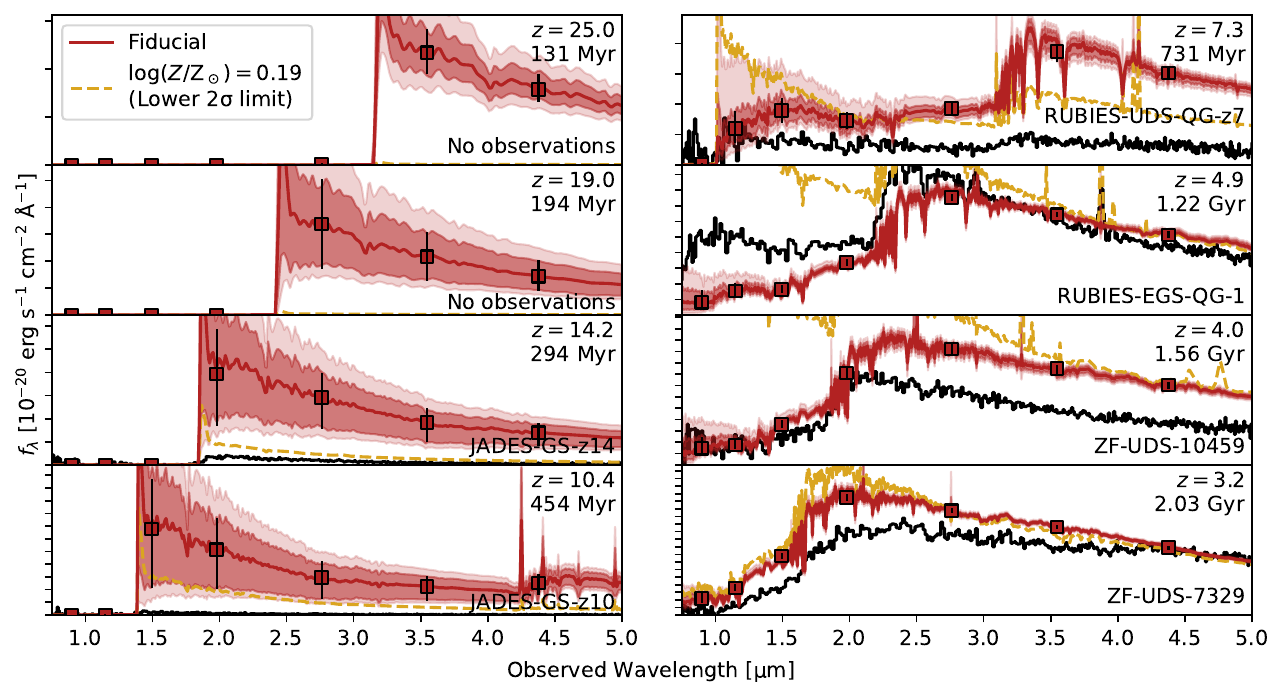}
\caption{The inferred posterior SED of \galaxy's progenitor (dark red shows $1\sigma$ uncertainty, light red $2\sigma$). We also show the lower $2\sigma$ limit of the fixed metallicity model's SED as a dashed goldenrod line. Each tick mark along the y-axis corresponds to $10^{-20}\ \mathrm{erg\ s^{-1}\ cm^{-2}\ \AA^{-1}}$. We compare the progenitor with the PRISM spectra of other galaxies from the literature: JADES-GS-z14-0 \citep{Carniani2024}, JADES-GS-z10 \citep{Curtis-Lake2023}, RUBIES-UDS-QG-z7 \citep{Weibel2025}, RUBIES-EGS-QG-1 \citep{DeGraaff2024a}, ZF-UDS-10459 \citep{Nanayakkara2024, Nanayakkara2025}, and ZF-UDS-7329 \citep{Glazebrook2024, Nanayakkara2024, Nanayakkara2025}. While the SED of \galaxy's progenitor qualitatively agrees with the latter three sources, the two $z>10$ JADES sources are barely visible in their respective panels and even the later-forming fixed metallicity model's SED is brighter than current high-$z$ observations.}
\label{fig:progenitor_spec}
\end{figure*}

In Figure~\ref{fig:progenitor_spec}, we show the inferred progenitor's SED posterior for \galaxy\ from the fiducial model. As a lower limit, we also show the lower $2\sigma$ flux of the fixed metallicity model's SED as a dashed goldenrod line. We show these SEDs at several redshifts where it would be classified as a ``dropout'' and where possible, we compare with PRISM spectra from other prominent high-redshift or quiescent galaxies in the literature: $z=25$ (no observations; F277W dropout), $z=19$ (no observations; F200W dropout), $z=14.2$ (JADES-GS-z14-0 \citealt{Carniani2024}; F150W dropout), $z=10.4$ (JADES-GS-z10 \citealt{Curtis-Lake2023}; F115W dropout), $z=7.3$ (RUBIES-UDS-QGz7 \citealt{Weibel2025}; F090W dropout), $z=4.9$ (RUBIES-EGS-QG1 \citealt{DeGraaff2024a}/UMG-28740 \citealt{UrbanoStawinski2024a}), $z=4$ (ZF-UDS-7542 \citealt{Nanayakkara2024, Nanayakkara2025}), $z=3.2$ (ZF-UDS-7329 \citealt{Glazebrook2024, Nanayakkara2024, Nanayakkara2025}). The spectra shown here are taken from the DJA and \emph{not} scaled to observed photometry. This means that the spectra of extended low-redshift sources may suffer from significant slit losses depending on NIRSpec shutter placement and source compactness \citep[see Fig. 5 and the discussion on spectrophotometric calibration in][]{Nanayakkara2024} in addition to typical wavelength-dependent slit losses. High redshift sources such as the $z>10$ JADES galaxies are generally compact \citep[][]{Robertson2024, Harikane2025} and are assumed to be ``high priority'' targets for spectroscopic followup and so would obtain favorable slit positions. Therefore we assume slit losses at $z>10$ are negligible enough for us to make qualitative comparisons. Figure~\ref{fig:progenitor_spec} shows that in every case, spectra of the brightest $z>10$ sources are dwarfed by the modeled progenitor's SED.

\begin{figure*}[!htb]
\centering
\includegraphics[width=\linewidth]{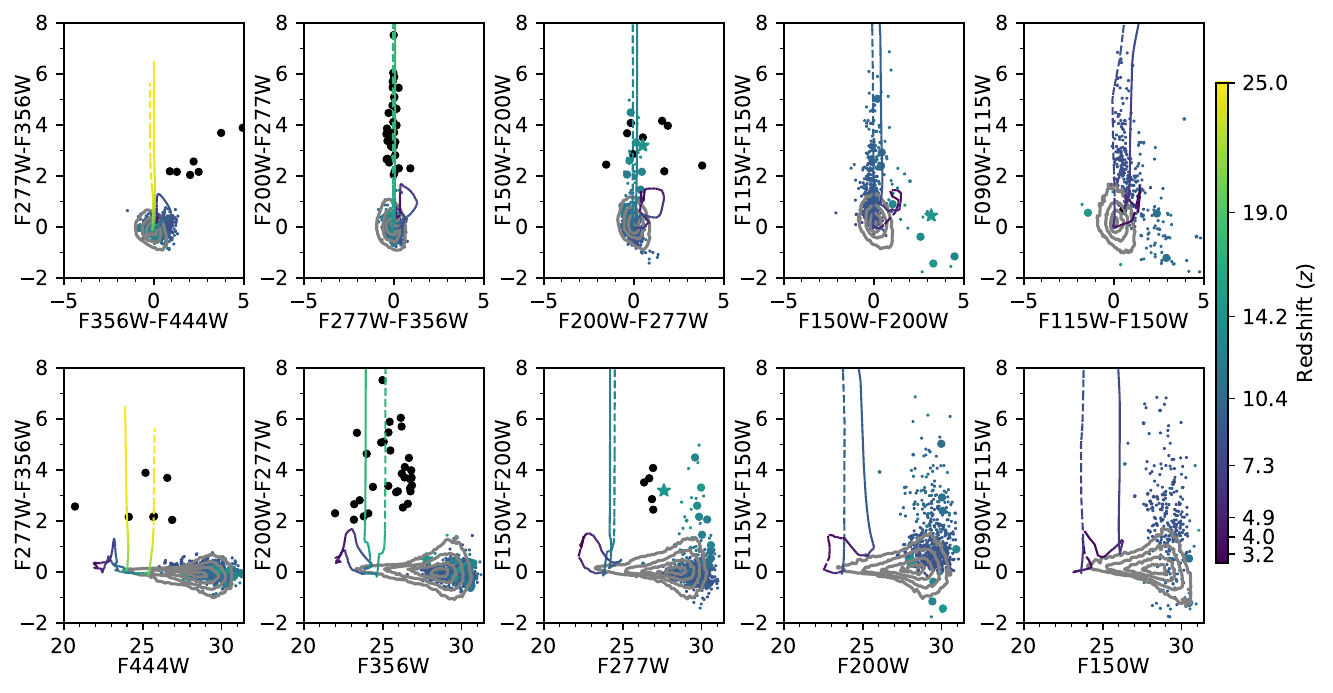}
\caption{\galaxy's inferred evolution through observed color-color (upper) and color-magnitude (lower) space in select broadband JWST/NIRCam filters. We show the median colors and magnitudes of the fiducial model and the fixed metallicity model by the solid line and dashed line, respectively, colored by redshift. We show high-redshift candidates from JADES (a star for JADES-GS-z14-0 from \citealt{Carniani2024}, large dots for $z>11$ from \citealt{Robertson2024}, and small dots for $z>8$ from \citealt{Hainline2024}). The contours show the distribution of all galaxies in the JADES catalog (GOODS-S) above the $5\sigma$ detection limit in \citep{Robertson2024}. ``Bright dropouts'' are indicated by black dots. All of these sources are ruled out as potential progenitor matches. }
\label{fig:progenitor_phot}
\end{figure*}

In the bottom panels of Figure~\ref{fig:progenitor_phot}, we show the posterior median of \galaxy's track through observed color-color and color-magnitude space for select JWST NIRCam broadbands as a solid colored line (we show the posterior median for the latest-forming model, fixed metallicity, as a dashed colored line). We also show the corresponding observed magnitudes and colors for high-redshift candidates from JADES (a star is used for JADES-GS-z14-0 from \citealt{Carniani2024}, large dots for $z>11$ from \citealt{Robertson2024}, and small dots for $z>8$ from \citealt{Hainline2024}). The inferred magnitudes of \galaxy's progenitor are clearly brighter than any previously identified ultra-high-redshift candidates. 

To investigate potential \galaxy-progenitor-like photometric candidates, for each color-color and color-magnitude panel pair (top and bottom), we show ``bright dropouts'' (defined as sources for which $\mathrm{F[bluest]W - F[middle]W > 2}$ and $\mathrm{F[reddest]W < 27}$) from the JADES catalog in GOODS-S. Of these sources, we select and inspect those which are fainter than 28.5 magnitudes in all bluer NIRCam or HST bands. Four F277W bright dropouts meet this criteria:
\begin{itemize}
    \item ID 92415 (RA 53.1337551, Dec -27.8255218),
    \item ID 140057 (RA 53.2003223, Dec -27.7641172,  $\mathrm{S/N}>10$ detection in F115W),
    \item ID 190413 (RA 53.0840401, Dec -27.8393486),
    \item and ID 203749 (RA 53.1214198, Dec -27.7949123; $\mathrm{S/N}>5$ detection in F200W).
\end{itemize}None of these sources have spectra on the DJA or entries in the spectroscopic FRESCO catalogs \citep{Meyer2024, Covelo-Paz2025} and all are HST-dark, appear compact, and have extremely red observed colors (see the top left panel in Figure~\ref{fig:progenitor_phot}), more consistent with AGN or Little Red Dots than with high-redshift Lyman breaks. All of the F200W or F150W bright dropouts have bright detections in at least one bluer band and therefore the observed colors cannot be due to high-redshift Lyman breaks. Even with this rudimentary selection, we can conclude that no sources in the GOODS-S field have colors consistent with the inferred unobscured progenitor of \galaxy. We perform the same search using the JADES GOODS-N catalog and again find zero candidates. 

Not only is the progenitor of \galaxy\ constructed from the properties and SFH inferred from our models significantly brighter than the already bright spectroscopically confirmed $z>10$ galaxy population, even rudimentary color and magnitude selections rule out all sources in both GOODS fields with JWST photometry. This could mean that that massive early galaxies are exceedingly rare or are significantly obscured by dust, or it implies singular early and fast-forming galaxies do not exist and \galaxy\ instead formed at later cosmic times or through mergers of several smaller galaxies.

\section{Bagpipes Modeling}
\label{sec:pipes}

Because our \texttt{Prospector} fits infer such early formation, we check our results with \texttt{Bagpipes}. Like \texttt{Prospector}, \texttt{Bagpipes} is a Python Bayesian fitting tool which generates model galaxy spectra for comparison with observed photometry and spectroscopy to infer observed galaxy properties. \texttt{Bagpipes} also allows for a variety of user-specified models and uses a Kroupa IMF \citep{Kroupa2001}. We opt to use the neural network-based \texttt{Nautilus} nested sampler \citep{Nautilus}.

\subsection{The Model}
We apply a \texttt{Bagpipes} model based on the one used to fit the quiescent galaxies presented in \citet{Carnall2024}. Metallicity is fit as a factor of solar metallicity with a logarithmic prior $Z/Z_\odot\in[0.00355, 3.55]$. We employ a \citet{Salim2018} dust law with $V$-band attenuation $A_V\in[0,4]$, deviation from the \citet{Calzetti2000a} slope $\delta\in[0.3, 0.3]$ (gaussian prior, $\mu=0,\ \sigma=0.1$), and a variable strength of the 2175\AA\ bump $B\in[0,5]$. We include the \citet{Draine2007} dust emission model with $q_\mathrm{PAH}\in[0.5,4]$, $\gamma\in[10^{-4},0.15]$ (logarithmic), and $U_\mathrm{min}\in[0.1, 15]$ (logarithmic). To account for the velocity dispersion of stars and gas, we include a gaussian convolution of the model spectrum with $\sigma\in[1,1000]\ \mathrm{km\ s^{-1}}$ (logarithmic). We also incorporate a noise jitter term on the spectrum $j_\mathrm{spec}\in[1,10]$ (logarithmic). The ionization parameter is fixed to $\log U=-0.3$. As with our \texttt{Prospector} modeling, we use a 10-degree polynomial to scale the spectrum to the photometry.

We fit three different versions of this model with differing SFHs. The first model uses a parametric double-power-law star formation history \citep[as in e.g.,][]{Carnall2017, Carnall2022b, Carnall2023a, Carnall2024, UrbanoStawinski2024a} with rising slope $\alpha$ and falling slope $\beta$ (both logarithmic priors in $[0.01, 1000]$) and a turnover time $\tau\in[0.1\ \rm{Gyr}, t_{\rm{obs}}]$ (logarithmic). For the parametric SFH model, we allow the redshift to vary $z_{\rm{spec}}\in[z_\mathrm{spec}-0.05, z_\mathrm{spec}+0.05]$.
 The second model assumes a nonparametric continuity SFH\footnote{Experimentation with other nonparametric SFHs using flexible bin widths \citep{Iyer2019} produces results consistent with those of the continuity SFH (extremely early formation times), so we opt to use the continuity SFH here to enable more direct comparisons with our earlier \texttt{Prospector} modeling.} with the same binning and priors as our fiducial \texttt{Prospector} model. For this model, we fix $z_\mathrm{spec} = 2.676$, inferred from fitting the first model. Finally, following the example of \citet{Carnall2024}, we fit a single instantaneous burst of star formation to obtain a lower limit of the formation time (see Appendix~\ref{sec:ssp} for comparisons with \texttt{Prospector} SSPs and the sources from \citealt{Carnall2024}).

\subsection{Analysis}

We tested our \texttt{Bagpipes} modeling using a \citet{Kriek2013} dust law (i.e., scaling the strength of the 2175\AA\ bump to the deviation from the \citealt{Calzetti2000a} slope $\delta$ rather than leaving them independent), but \galaxy\ is faint in the UV and the spectral coverage does not extend to rest-frame 2175\AA\ to constrain the bump, so this choice has no effect on the inferred stellar population properties or SFHs. We tested using uniform rather than gaussian or logarithmic priors on the dust parameters and $j_\mathrm{spec}$ and found our fits robust to how these priors were scaled.

In our \texttt{Bagpipes} fits, we mask out the same regions as we do in our \texttt{Prospector} modeling. We additionally mask out the region around the H$\alpha$ and \NII\ features in our \texttt{Bagpipes} fits as in \citet{Carnall2024} because the code can only model emission lines from ongoing star formation. Our results are robust to whether or not this region is masked.

\subsection{Results}

\begin{figure*}[!htb]
\centering
\includegraphics[width=\linewidth]{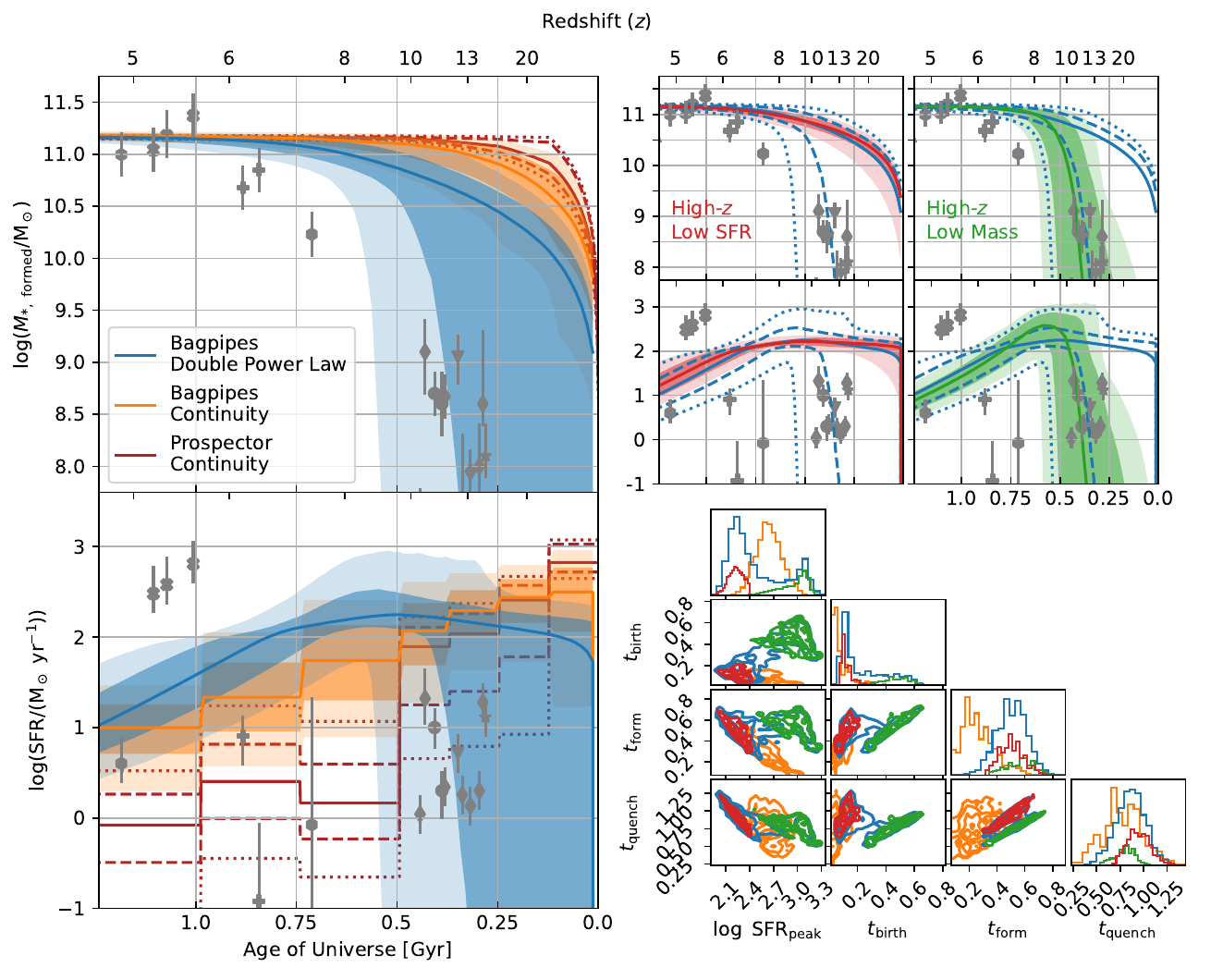}
\caption{\textbf{Left:} the SFH posteriors of \galaxy\ modeled with \texttt{Bagpipes}, as Figure~\ref{fig:sfh_fiducial} (here we have rescaled age of the universe and SFR axes to focus on the early universe). Blue shows the parametric double power law SFH and orange is the continuity SFH (the \texttt{Prospector} fiducial model's median/$1\sigma$/$2\sigma$ is shown by the solid/dashed/dotted red lines). \textbf{Top right:} posteriors of the subset of SFHs with $\mathrm{SFR}<250\ \mathrm{M_\odot\ yr^{-1}}$ at $z>10$ (left, red) and $\log (M_\mathrm{*,\ formed}/\mathrm{M_\odot})<9.5$ at $z>15$ (right, green), shown with the same axes as the panels on the left. In the plots with the subpopulations, we show the entire double power law SFH posterior median/$1\sigma$/$2\sigma$ by solid/dashed/dotted blue lines. \textbf{Bottom right: }a corner plot of the maximum SFR and the time in Gyr at which each SFH would have assembled $10\%$, $50\%$, and $90\%$ of its stellar mass ($\log\ \mathrm{SFR_{peak}}$, $t_\mathrm{birth}$, $t_\mathrm{form}$, and $t_\mathrm{quench}$ respectively) for the nonparametric continuity (orange) and parametric double power law models (blue). We also show the high redshift low SFR (red) and high redshift low stellar mass (green) subpopulations of the parametric model which trace out the bimodal trend of either late and rapid or early and slow mass assembly.}
\label{fig:pipessfh}
\end{figure*}

The properties of \galaxy\ inferred with \texttt{Bagpipes} are in good agreement with those from our \texttt{Prospector} modeling. In Table~\ref{tab:pipetable} we show the surviving stellar mass and inferred formation/quenching times/redshifts using the same definition as in \S\ref{sec:p_analysis} (the fiducial \texttt{Prospector} model is included for comparison). The \texttt{Bagpipes} continuity SFH model infers somewhat later formation times than our \texttt{Prospector} modeling (though still very early). The double power law model infers slightly later formation and quenching than either continuity model with much larger uncertainties in SFR and mass at high redshift. In the left panels of Figure~\ref{fig:pipessfh}, we show a high-redshift zoom-in on the SFH and mass assembly history of our double power law and continuity SFH \texttt{Bagpipes} fits (we show $\log M_{*,\ \mathrm{formed}}/\mathrm{M_\odot}$ in this figure rather than $\log M_{*}/\mathrm{M_\odot}$). The median/$1\sigma$/$2\sigma$ uncertainties of the fiducial \texttt{Prospector} model are shown by the solid/dashed/dotted red lines. 

\begin{deluxetable*}{cccccccc}
\centerwidetable
\tablewidth{0pt}
\tablecaption{Mass and Formation History of \galaxy\ inferred by \texttt{Bagpipes} SED modeling \label{tab:pipetable}}
\tablehead{\colhead{Model} & \colhead{$z_\mathrm{spec}$} & \colhead{$\log M_*/\mathrm{M_\odot}$} & \colhead{$t_\mathrm{form}$} & \colhead{$z_\mathrm{form}$} & \colhead{$t_\mathrm{quench}$} & \colhead{$z_\mathrm{quench}$}\\
\colhead{} & \colhead{} & \colhead{} & \colhead{[Myr]} & \colhead{} & \colhead{[Myr]} & \colhead{}
}
\startdata
Double Power Law	&	$2.6760_{-0.0002}^{+0.0002}$	&	$10.91_{-0.01}^{+0.02}$	&	$496^{+110}_{-104}$	&	$9.55_{-1.32}^{+1.81}$	&	$856^{+158}_{-155}$	&	$6.33_{-0.79}^{+1.05}$\\
Continuity	&	$2.6760$	&	$10.94_{-0.02}^{+0.02}$	&	$242^{+136}_{-91}$	&	$16.00_{-4.38}^{+6.32}$	&	$750^{+195}_{-187}$	&	$7.00_{-1.15}^{+1.69}$\\
Burst$^{a}$ &	$2.6760$  &	$10.92^{+0.01}_{-0.01}$	&	$480^{+158}_{-164}$	&	$9.98_{-1.90}^{+3.53}$&	---	&	--- \\
\hline
Fiducial (\texttt{Prospector})	&	$2.6748_{-0.0002}^{+0.0001}$	&	$10.96_{-0.01}^{+0.01}$	&	$108_{-39}^{+37}$	&	$28.41_{-5.29}^{+10.31}$	&	$353_{-179}^{+53}$	&	$12.46_{-1.20}^{+8.10}$\\
\enddata
\tablecomments{The inferred formation and quenching times/redshifts for \galaxy\ from \texttt{Bagpipes} modeling, in the same format as Table~\ref{tab:megatable2}. We include the fiducial \texttt{Prospector} model here for comparison.}
\tablenotetext{a}{The burst SFH is a SSP with a single age, so we give $t_\mathrm{form}=t_\mathrm{obs}-t_\mathrm{age}$. See also Appendix~\ref{sec:ssp}.}
\end{deluxetable*}

\subsubsection{Nonparametric SFHs in Early Quiescent Galaxies}
\label{sssec:nonparaprobs}

In comparing parametric and nonparametric models, \citet{Carnall2019a, Leja2019} arrived at the conclusion that nonparametric SFHs (especially the continuity SFH) typically outperform parametric SFHs at recovering the ``true'' SFH of $z\sim1$ low-redshift quiescent galaxies. However, we find here that the nonparametric continuity SFH infers disconcertingly early formation for this high-redshift quiescent galaxy, even with physically-motivated priors  (these results may be exacerbated by the difficulty of reliably and correctly inferring the global likelihood maximum; see Appendix \ref{sec:sampling} and \citealt{Wang2025}). This is similarly the case for ZF-UDS-7329, which was fit with both a double power law parametric SFH and a nonparametric continuity SFH in \citet{Glazebrook2024}. In their Figure 3, the \texttt{Prospector} nonparametric SFH is already at peak star formation at $z>100$ when the \texttt{FAST++} parametric SFH is still rising \citep[also compare with the other quiescent galaxies in][]{Nanayakkara2024, Nanayakkara2025}. 

While the extremely early formation times may be partially driven by the \texttt{Prospector} stellar libraries and isochrones preferring an older stellar population than \texttt{Bagpipes} (see Appendix \ref{sec:ssp}), we find that in the \texttt{Bagpipes} continuity model the highest SFRs occur in the earliest bin as well. We speculate that could reflect a systematic bias in extremely early-forming (quiescent) galaxy SFHs inferred with nonparametric models. In the continuity model, a prior is applied to the ratio between neighboring bins, therefore bin $i$'s SFR is weighted towards that of bin $i-1$ and $i+1$. However, the first bin and last bin only have a single neighboring bin and are thus less penalized for extreme changes in SFR. This can be intuited by considering the prior on $\log \mathrm{SFR}_i/\mathrm{SFR}_{i+1}$: for any set of $N$ bins, each bin is subject to two priors \emph{except} bins $i=1$ and $i=N$. Therefore, the earliest bin allows for burstier star formation which in turn results in unphysically rapid and early inferred galaxy formation and quenching, as we see with \galaxy. This ``first-bin-burstiness problem'' may therefore introduce another unstated and unexplored systematic in the models of extremely early quiescent galaxies.

We further speculate that the most recent SFH bin is similarly subject to this trend \citep[see also][]{Wang2025} and may exacerbate ``outshining'' in these populations, \citep[e.g.,][]{Papovich2001, Narayanan2024}, though the implications are less cosmically problematic as lower-redshift extreme star formation is more ``reasonable'' than equally extreme star formation at $z>20$. Nevertheless, we encourage caution when interpreting SFHs with extreme star formation in either the most recent or earliest bin. Future development and experimentation comparing SFH models with mock galaxies informed by new high-redshift galaxy formation models will be necessary to address this potential problem.

\subsubsection{Parametric SFHs in Early Quiescent Galaxies}

At first glance, it appears that the double power law SFH is less at odds with observations. Are SFHs inferred with parametric double power laws better able to match high-redshift observations? 

To investigate the trends in the inferred parametric SFH, we select from the modeled SFHs two sub-populations, the set of all model SFHs for which $\mathrm{SFR} < 250\ \mathrm{M_\odot\ yr^{-1}}$  at $z>10$ and the set of all model SFHs which have assembled $\log M_{*,\ formed}/\mathrm{M_\odot}<9.5$ by $z=15$ (high-$z$ low-SFR and high-$z$ low-mass, shown by the red and green SFHs in Figure~\ref{fig:pipessfh}, respectively). Out of 5000 SFHs sampled from the posterior, no single SFH fits both criteria. In the bottom right of Figure~\ref{fig:pipessfh} we show a corner plot with the inferred posteriors of the maximum SFR and the time at which the model would have assembled $10\%$, $50\%$, and $90\%$ of its stellar mass ($\log\ \mathrm{SFR_{peak}}$, $t_\mathrm{birth}$, $t_\mathrm{form}$, and $t_\mathrm{quench}$ respectively).

Clearly the parametric SFH posterior inferred with \texttt{Bagpipes} contains two distinct modes: an extreme short burst of star formation at later times or a slow buildup of stellar mass which always starts early. This bimodality also appears to trend with the previously selected high-$z$ low-mass and high-$z$ low-SFR SFH subsets. In the first category, SFR jumps to $\gtrsim100\ \mathrm{M_\odot yr^{-1}}$ and remains constant for the first $\sim\mathrm{Gyr}$. In this mode, the model quickly assembles $10\%$ of its stellar mass but does not cross the $50\%$ and $90\%$ thresholds until much later (lower branch in $t_\mathrm{birth}-t_\mathrm{form}$ and left hand side blob in the $t-\log\ \mathrm{SFR_{peak}}$ covariance panels). In the second category, SFR initially remains low until $z\sim10-15$ but then promptly increases to $\approx500\ \mathrm{M_\odot yr^{-1}}$ and stellar mass is rapidly assembled on a short timescale. Here, $t_\mathrm{birth}$, $t_\mathrm{form}$, and $t_\mathrm{quench}$ are all highly correlated (diagonal path in all $t$-versus-$t$ panels). Furthermore, we can see from the $t_\mathrm{form}$ and $\log\ \mathrm{SFR_{peak}}$ that the later peak SFRs tend to be more extreme than individual galaxies from high redshift observations ($\mathrm{SFR_{peak}}\approx500-1000\ \mathrm{M_\odot yr^{{-1}}}$ for $z_\mathrm{form}\sim10$).

The first mode, early and steady star formation (e.g., the red posterior), is still more massive and more star forming earlier than high-redshift observations. The second mode, which allows later bursts, qualitatively appears to match high-redshift observations (the observed points lie within the green posterior in Figure~\ref{fig:pipessfh}), however at this point in time the inferred SFH is \emph{rapidly} rising. Such strongly star-forming galaxies have not been observed at $z\gtrsim10$. In other words, the modes hidden within the double power law SFH posterior still do not match high redshift observations, they are either too massive or too star-forming.

Some individual models' SFH are both late and gradual, but these are disfavored as the inferred photometry for these models is too bright blueward of the $4000\ \mathrm{\AA}$ break and too faint redward, an indication that the model's stellar population is too young (similar to the fixed metallicity model in Figure~\ref{fig:phot_model}). However, all of the modeling done thus far has assumed that elemental abundance ratios are scaled to solar values. In \citet{Park2024}, the new $\alpha$-enhanced stellar models have notably different SED shapes than typical solar-scaled models from $0.4-1.0\ \mu$m, the region where our modeling has difficulty reproducing the observed photometry (i.e., Figure~\ref{fig:phot_model}). In their reanalysis of the PRISM spectrum for ZF-UDS-7329, \citet{Turner2025} additionally showed that an $\alpha$-enhanced SSP can mimic the colors of an older $Z=\mathrm{Z_\odot}$ stellar population. Therefore, $\alpha$-enhancement offers a potential explanation for the early and rapid inferred formation of \galaxy\ and the difficulty matching observed the observed SED in the redder NIRCam bands.

\section{$\alpha$-enhancement}\label{sec:alfalpha}

The chemical evolution of stellar populations is highly sensitive to formation timescales. Because $\alpha$ elements like Mg and Ca are produced when massive short-lived stars undergo core-collapse supernovae (CC SNe), a galaxy's ISM (and subsequently formed stars) experience early enrichment by these elements. Other elements, such as Fe, enter the ISM through other pathways: Fe is produced in equal amounts in CC SNe and Type 1a supernovae (SNe Ia). Whereas enrichment from CC SNe is close to instantaneous on the timescales discussed here, SNe Ia only ``turn on'' after $\sim0.5-1$ Gyr \citep{Maoz2010}. Therefore, galaxies which form the bulk of their stars before SNe Ia can enrich their ISM exhibit higher abundances of $\alpha$ elements than e.g. Fe \citep{Kriek2016a, Beverage2024, Beverage2025, Carnall2024, Carnall2022b, Gountanis2025}.

We fit the observed spectrum with \texttt{alf$\alpha$} \citep{Beverage2025} to investigate whether \galaxy\ is $\alpha$-enhanced. \texttt{alf$\alpha$} is a Python SED-modeling code based on the Absorption Line Fitter \citep[\texttt{alf};][]{Conroy2018}, which infers elemental abundances from stellar spectra for sufficienctly old ($\gtrsim1\ \mathrm{Gyr}$) stellar populations. \texttt{alf$\alpha$} uses a Kroupa IMF \citep{Kroupa2001} and we run \texttt{alf$\alpha$} with the nested sampling code \texttt{dynesty}.

\subsection{The Model}

  \texttt{alf$\alpha$} fits observed spectra (no photometry) to SSP models assembled from the empirical MILES and IRTF spectral libraries, and metallicity-dependent MIST isochrones. To probe the abundances of individual elements, theoretical response spectra for 19 elements are included in these models. In the fitting, we apply a 7 degree Chebyshev polynomial to scale the spectrum to the SSP models.  

 We use the default priors for [X / H] for the elemental response functions. For Fe, C, N, O, Mg, Si, Ca, Ti, and Cr, [X / H] $\in$ [-0.5, 0.5]. Two additional ratios have specific priors: [Na / H] $\in$ [-0.5, 1.0] and [Z / H] $\in$ [-1.5, 0.3]. Though our primary goal is to infer [Mg/Fe] to determine if \galaxy\ is $\alpha$-enhanced, we include the whole suite of elemental response functions and treat all of the others as nuisance parameters. We also fit for $\sigma_{*}$, $T_{\rm{eff}}$, a spectroscopic jitter term, $\sigma_{\rm{gas}}$, and relative velocity offset of Balmer, \OIII, \NII, and \SII\ emission lines (all using default priors). Finally we set a uniform prior on log age $\in$ [-0.3, $\log t_{\rm{univ}}$].

\subsection{Analysis} \label{ssec:alfalfa}

Our earlier SED modeling of \galaxy\ with \texttt{Prospector} and \texttt{Bagpipes} infers early and rapid formation, therefore we expect it would be $\alpha$-enhanced. Due to the inferred old age of \galaxy ($\gtrsim2\ \mathrm{Gyr}$), we are able to acquire individual elemental abundances from the spectrum with \texttt{alf$\alpha$}.

\begin{figure*}[!htb]
\centering
\includegraphics[width=\linewidth]{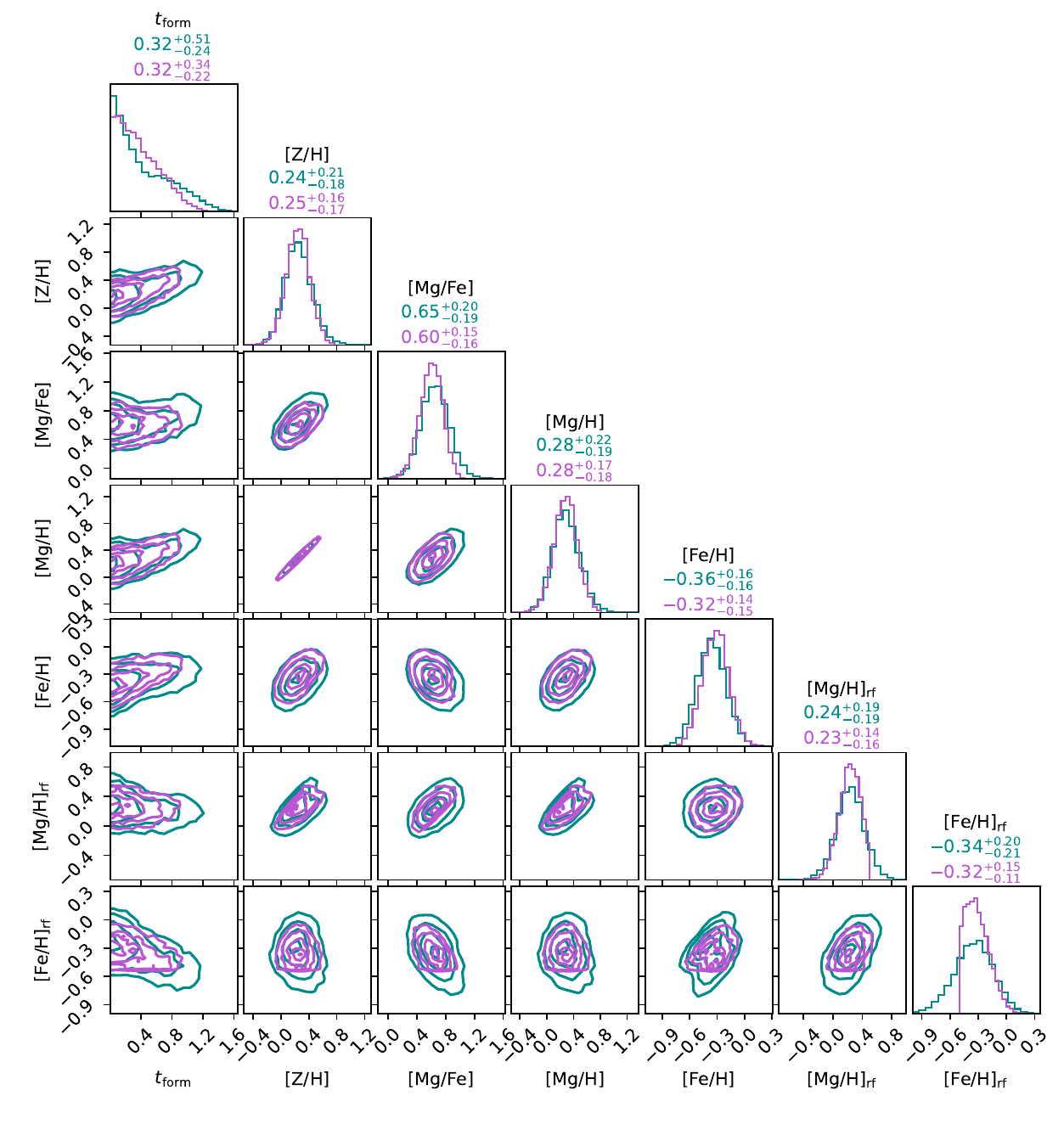}
\caption{A corner plot of a subset of inferred quantities from \texttt{alf$\alpha$} spectral modeling. We show results from runs with both the default priors (purple) and widened response function priors (dark cyan). When the default priors are used for the elemental abundance response functions, $\mathrm{[Mg/H]_{rf}}$ and $\mathrm{[Fe/H]_{rf}}$ can be seen to run up against the edges of the priors. The total abundances [Mg/H] and [Fe/H] are calculated by combining the abundances inferred from the response functions with those from the solar scaled model SSP. Both models prefer early $t_{\rm{form}}$ of the SSP, but some models form as late as 1 Gyr after the Big Bang ($z\sim6$).
}
\label{fig:alf_corner}
\end{figure*}

However first we caution that due to the relatively low S/N of the spectrum ($\sim7.5$ per rest-frame \AA), the majority of inferred [X/H] abundances are not be robust. We only report results for elements with strong stellar features, e.g., Mg (Mg I) and Fe (G-band, Fe I). Because of the relatively low S/N of our spectrum, we find that the inferred abundances suffer from large uncertainties and in some instances the posteriors run up against the edge of the prior (see the magenta contours and histograms for the abundances in the response functions, [Fe/H]$_{\rm{fn}}$ and [Mg/H]$_{\rm{fn}}$ in Figure \ref{fig:alf_corner}). Therefore, we run \texttt{alf$\alpha$} a second time and widen every [X/H] prior by 0.5 dex in each direction to better infer the uncertainties of the abundances.

Figure \ref{fig:alf_corner} shows the corner plot for abundances and properties measured by \texttt{alf$\alpha$} with default (purple) and widened (dark cyan) abundance priors. The response functions are denoted as $\mathrm{[X/H]_{rf}}$ (note the subscript) and differ from the total abundances $\mathrm{[X/H]}$ in that the latter is calculated using both the response function and the abundances built into the solar-scaled SSP. The abundances and elemental ratios generally agree between the default prior and widened prior fits, but the response function abundances inferred with default priors run up against the prior limits. Therefore, we take the values from the fit with widened priors as the fiducial values.

\begin{figure}[!htb]
\centering
\includegraphics[width=\linewidth]{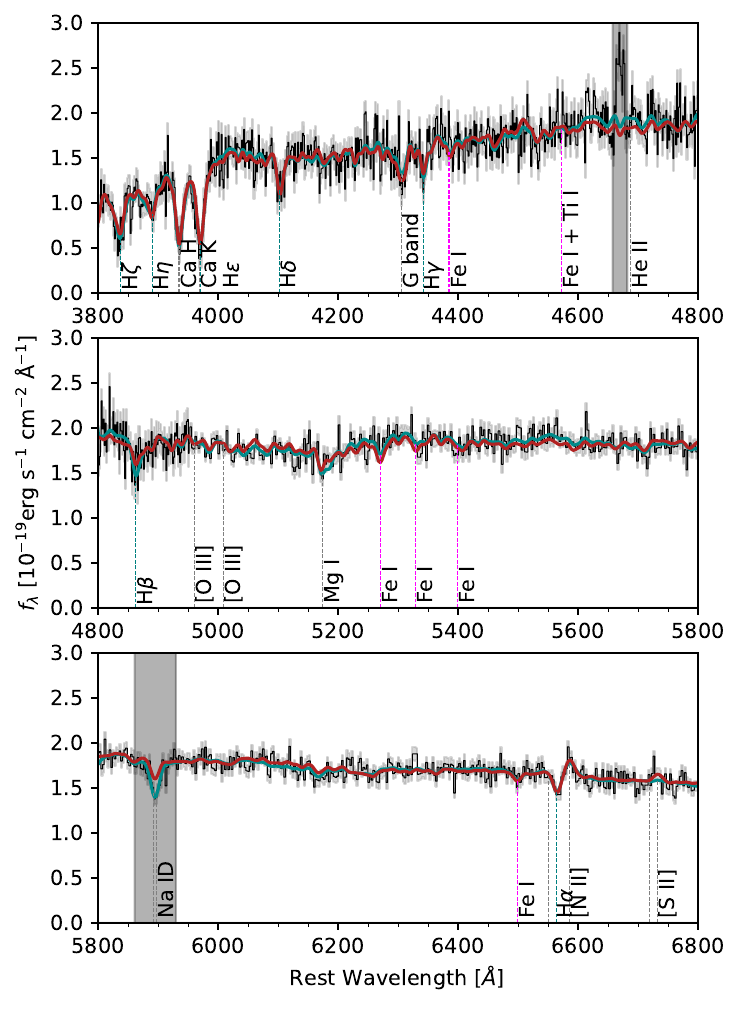}
\caption{The observed combined spectrum (black), associated uncertainty (gray, along the bottom), best-fit \texttt{Prospector} model spectrum (red), and best-fit \texttt{alf$\alpha$} model (dark cyan). We mark the locations of common emission/absorption features with dashed lines and label them. While the \texttt{alf$\alpha$} model generally fits the spectrum well, the S/N ($\sim7.5$ per rest-frame \AA) is insufficient for robust abundance measurements of all but the most significant absorption features. Although the Na ID absorption feature is masked out in all our fits and is not matched by the MAP \texttt{Prospector} model, the \texttt{alf$\alpha$} model reproduces it excellently \citep[we also detect no evidence of outflows unlike some other quiescent galaxies, e.g.][]{Belli2024a, Davies2024b}. As opposed to Figure \ref{fig:spec_model}, here the spectrum is uncalibrated and we apply the polynomial to the models to match them to the observed spectrum.
}
\label{fig:spec_zoom}
\end{figure}

\subsection{Results}

From our \texttt{alf$\alpha$} fitting, we infer $\mathrm{[Fe/H]}=-0.36^{+0.16}_{-0.15}$ and $\mathrm{[Mg/H]}=+0.29^{+0.22}_{-0.19}$. These results indicate that \galaxy\ is deficient in Fe\footnote{Like the Bronze Age city, get it?} relative to quiescent galaxies at lower redshifts \citep[$z\sim0$ and $z\sim0.7$][]{Beverage2023}, and has comparable [Mg/H]. The JWST SUSPENSE survey \citep{Slob2024}, which acquired abundances for a sample of high S/N $1<z<3$ quiescent galaxies also generally found their sample to be deficient in Fe and C with [Mg/H] slightly below $z<1$ quiescent population \citep{Beverage2025}. 

From the inferred [Mg/H] and [Fe/H] we calculate $\mathrm{[Mg/Fe]}=+0.65^{+0.20}_{-0.19}$, which indicates that \galaxy\ is significantly $\alpha$-enhanced, as is expected for massive galaxies which quenched quickly at high redshift \citep{Kimmig2025}. Though the uncertainties are large, this value is consistent with other early-forming $\alpha$-enhanced galaxies, such as ZF-UDS-7329 at $z=3.2$ \citep[$\mathrm{[Mg/Fe]}=+0.42^{+0.19}_{-0.17}$;][]{Carnall2024}, the lensed quiescent galaxy at $z=1.98$ \citep[$\mathrm{[Mg/Fe]}=+0.51\pm0.05$;][]{Jafariyazani2020}, the highest [Mg/Fe] in the SUSPENSE survey \citep[ID 129133 with $\mathrm{[Mg/Fe]}=+0.51^{+0.07}_{-0.08}$, $z=2.139$, $\log M_*/M_{\odot}=11.1$;][]{Beverage2025}. The [Mg/Fe] ratio of \galaxy\ is most comparable to that of the quiescent $z=2.1$ galaxy from \citet{Kriek2016a} with $\mathrm{[Mg/Fe]}=+0.59\pm0.11$, which was shown to be consistent with a star-formation timescale of $0.1-0.5$ Gyr. For \galaxy, this formation timescale would require star formation on the order of $200-1000\ M_{\odot}\ \mathrm{yr^{-1}}$, which aligns with the peak SFRs from the inferred SFHs. Although \texttt{alf$\alpha$} prefers maximal ages for \galaxy with the mode of $t_{\mathrm{form}}$ (where $t_\mathrm{form}=t_\mathrm{obs}-t_\mathrm{age}$) falling at the Big Bang, it does allow formation times as late as $\lesssim1.5$ Gyr (or a lookback time of $\gtrsim1$ Gyr). Whereas \citet{Beverage2025} found that the formation timescales for the SUSPENSE galaxies from \texttt{alf$\alpha$} conflicted with the reported nonparametric \texttt{Prospector} SFHs presented in \citet{Slob2024}, we find that in \galaxy\ the preferred early formation time and fast star-formation timescale from our inferred $\alpha$-enhancement agree with the \texttt{Prospector} modeling.

Using the formulation $\log (Z/\mathrm{Z_{\odot}}) = \mathrm{[Z/H]} = \mathrm{[Fe/H]} + 0.94\times\mathrm{[Mg/Fe]}$ from \citet{Thomas2003}, we calculate a metallicity  $\log Z/Z_{\odot} = +0.25^{+0.21}_{-0.18}$. Our inferred \texttt{Prospector} metallicity is $\sim0.4$ dex lower than this derived \texttt{alf$\alpha$} metallicity, though it seems to agree better with the \texttt{alf$\alpha$} [Fe/H], as was also true for the SUSPENSE sample \citep{Beverage2025}.

 We remind the reader of the significantly widened elemental response function priors, modified so because of the relatively low S/N of \galaxy's spectrum. \citet{Conroy2018} initially computed these response functions in the range $\mathrm{[X/H]\in[-0.3,+0.3]}$ in most cases. While linear combinations of individual response functions with relatively modest shifts in [X/H] produce only minor changes in the SSP spectrum, significant departures from the base abundance patterns in the SSP may be unrealistic. However, while the uncertainties of $\mathrm{[X/H]_{rf}}$ from our fits are large, the posterior medians are well within the base $[-0.5, +0.5]$ range set in \texttt{alf$\alpha$} (see Figure \ref{fig:alf_corner}) and the widened-priors best-fit \texttt{alf$\alpha$} SSP is at least qualitatively similar to the best-fit \texttt{Prospector} fiducial model and the observed spectrum. In other words, the upper/lower limits of enhanced/deficient [X/H] abundances should be treated with some skepticism, but the lower/upper limits and corresponding medians reported here are not immediate cause for concern.

While our \texttt{alf$\alpha$} analysis does not conclusively demonstrate that \galaxy\ formed extremely early and the uncertainties are large, the inferred [Mg/Fe] unambiguously indicates that it is $\alpha$-enhanced, signaling that \galaxy's formation was very rapid. This $\alpha$-enhancement is a plausible culprit for the disagreement between the observed photometry and models fit with \texttt{Prospector} and \texttt{Bagpipes} which utilize stellar libraries with solar-scaled stellar abundances. Much like how ZF-UDS-7329 at $z=3.2$ has been a powerful test case for stellar population and galaxy formation models, \galaxy\ is another early-forming test case for the next generation of sophisticated galaxy models. In the future, implementation of the non-solar scaled $\alpha$-MC stellar libraries \citep{Park2024} in FSPS will allow for more accurate SED modeling of these early-forming galaxies.

\section{Environment}
\label{sec:env}

Many other early-forming quiescent galaxies have been shown to exist in overdense regions \citep{Kubo2021b, McConachie2022a, Ito2023a, Tanaka2024a, DeGraaff2024a, UrbanoStawinski2024a, Jespersen2025, Kawinwanichakij2025} or near other massive galaxies \citep{Schreiber2018a, Carnall2024, Turner2025}. Observations of $z>10$ cosmic dawn galaxies in close proximity to each other also hint at environment's influence on the formation of the first galaxies \citep{Tacchella2023, Carniani2024}.

Perhaps then it should come as no surprise that \galaxy\ lies in close proximity to a massive protostructure as well. Using photometric redshifts and ground-based spectroscopy, \citet{Shah2024} mapped the GOODS-S field and uncovered \emph{Drishti}, a contiguous structure with $\log M_\mathrm{tot}/\mathrm{M_\odot}=14.9$ at $2.64<z<2.71$ consisting of three massive overdense peaks. The most massive and most overdense peak (P1\_S1, $\log M/\mathrm{M_\odot}=13.7$) lies at the same redshift as \galaxy\ with its center offset by 1.91 proper Mpc from the galaxy. the second and third most massive peak (P3\_S1 and P2\_S1, $\log M/\mathrm{M_\odot}=13.5$ and $13.4$) are at a slightly higher redshifts $z=2.697$ and $z=2.694$ but are also roughly equidistant from \galaxy\ on the sky. Extended structure around these peaks appears to extend into the region surrounding \galaxy\ \citep[Figures 2 and 3 in][]{Shah2024} with additional sub-peaks scattered throughout.

The fact that an ancient galaxy like \galaxy\ is found in a massive protostructure like \emph{Drishti} suggests a connection to environment. If \galaxy\ were situated in a more massive subhalo in the protostructure, it would allow for higher, earlier star formation rates than an otherwise isolated galaxy which in turn would allow for more extremely massive and early galaxy \citep[e.g., ][]{Jespersen2025}. Overdense environments also offer an alternative mode of stellar mass assembly via mergers with other neighboring galaxies (e.g., potentially the blue fluff to the southeast in the cutout in Figure~\ref{fig:spec_model}). If the stars seen in \galaxy\ at $z=2.675$ assembled across separate halos at high redshift (i.e., ex-situ) before merging into \galaxy\ at later times, it could explain the lack of high-redshift unobscured progenitor candidates. While major mergers at high redshift are rare, such an event could help explain the expected rapid and early quenching, as they have been proposed as quenching mechanisms \citep{DiMatteo2005}. Future work investigating the connection between \galaxy\ and \emph{Drishti} is therefore a crucial piece of the puzzle to unraveling the effects of environment on the formation and evolution of \galaxy\ and other early quiescent galaxies.

\section{Conclusions} \label{sec:conc}

In this work we presented \galaxylong\ (nicknamed ``\galaxy''), an ancient galaxy at $z=2.675$ with $\log M_*/\mathrm{M_\odot}=10.96^{+0.01}_{-0.01}$ which is inferred to have formed when the Universe was $<500$ Myr old. We summarize our conclusions below:

\begin{itemize}
    \item Observations of \galaxy\ in the GOODS-S field form an exquisite dataset. With imaging from legacy HST programs, extensive JWST/NIRCam campaigns, and a deep JWST/MIRI survey, photometric coverage of \galaxy\ extends from $4-25\ \mu$m. Its spectrum was measured from $1-3\ \mu$m with deep medium resolution JWST NIRSpec follow-up in the G140M and G235M grating by the SMILES program. The general SED shape from photometry and prominent spectral features such as a $4000$ \AA\ break, deep Ca H\&K lines, relatively weak H$\delta$, and other stellar absorption features like Mg I all indicate that \galaxy\ is already extremely mature when the universe is only $\sim2.5\ \mathrm{Gyr}$ old.
    
    \item We model \galaxy\ using the Bayesian inference Python package \texttt{Prospector} to simultaneously fit the photometry and combined G140M+G235M spectrum. Our model makes use of a nonparametric star formation history to infer when \galaxy\ formed and quenched. We further explore alterations to the fitting by changing input data, limiting the onset of star formation, applying physically motivated priors to the star formation history, and changing stellar population properties (IMF, metallicity). In almost every case, \texttt{Prospector} infers a SFH with an extremely early and rapid formation ($t_\mathrm{form}\lesssim300\ \mathrm{Myr}$) followed by early quenching ($t_\mathrm{quench}\lesssim500\ \mathrm{Myr}$). Only by forcing the model to adopt supersolar metallicites can the inferred SFH be slowed, albeit at the cost of a worse fit to observed photometry. All of the inferred SFHs for \galaxy\ are more star-forming and/or more massive than even the brightest and most star-forming high-redshift observations.

    \item Using the inferred properties and SFH of \galaxy, we model its potential progenitors and compare their spectral shapes and JWST NIRCam colors to observations. The spectra of \galaxy's unobscured progenitor dwarfs the brightest high-redshift spectroscopic observations and its modeled NIRCam fluxes are magnitudes brighter than any candidate yet identified in the GOODS fields. A search through the JADES GOODS-S and GOODS-N catalogs for other sources with similar colors and magnitudes finds zero candidates. 

    \item We check our \texttt{Prospector} modeling by also fitting \galaxy\ with \texttt{Bagpipes}. As with \texttt{Prospector}, we recover early formation over a rapid timescale. By assuming a parametric double power law SFH, the posterior better aligns with high redshift observations. Further investigation reveals that the parametric SFH posterior contains two modes: one which is too massive too early, and one which is too star-forming at later times. We discuss how the extremely high early star formation rates inferred for early quiescent galaxies with the nonparametric continuity SFH model may be a consequence of the prior under-penalizing large deviations in the SFR of the first and final bins, leading a ``first bin burstiness problem.''
    
    \item We infer abundances for iron and magnesium in \galaxy\ by fitting its spectrum with \texttt{alf$\alpha$}. We calculate $\mathrm{[Mg/Fe]}=+0.65^{+0.20}_{-0.19}$, indicating that \galaxy\ is strongly $\alpha$-enhanced, which supports the rapid formation timescales inferred by our \texttt{Prospector} and \texttt{Bagpipes} modeling. While the \texttt{alf$\alpha$} model also prefers a very old SSP, it does allow for formation as late as $z\sim6$ in some instances. We speculate that SED modeling with $\alpha$-enhanced stellar libraries could lessen \galaxy's tension with high-redshift observations if a younger $\alpha$-enhanced population is better able to model its observed SED.
    
    \item \galaxy\ lives in the massive ($M_\mathrm{tot}=10^{14.9}\mathrm{M_\odot}$) protostructure \emph{Drishti} at $2.64<z<2.71$ in GOODS-S \citep{Shah2024}. The high density environment of \emph{Drishti} provides additional mechanisms, e.g., mergers, which could explain fast early mass assembly and rapid quenching. That such an old galaxy is found within such a massive structure indicates that the environment of these early-forming monsters is likely a crucially important factor in the formation and evolution of such early-forming galaxies.

\end{itemize}

It is standard practice to infer the properties of galaxies with SED modeling tools. The inferred SFHs of some old/high-redshift massive quiescent galaxies like \galaxy, ZF-UDS-7329 \citep{Glazebrook2024}, PRIMER-EXCELS-109760 \citep{Carnall2024}, RUBIES-EGS-QG1 \citep{DeGraaff2024a} would imply the existence of $z\gtrsim10$ galaxies with $M_*\approx10^{10}-10^{11}\ \mathrm{M_\odot}$, which models and simulations have difficulty predicting. However, such massive high-redshift galaxies have not yet been found observationally either. Therefore, one must conclude that either the light from massive progenitors are heavily attenuated by dust, the stars found in the low-redshift quiescent descendant are formed ex-situ across many halos at high redshift, or the systematics of standard SED modeling tools used to fit galaxies are significantly underestimated (at least at high redshift).

Though the formation history of \galaxy\ inferred by SED modeling is confounding, it is undeniably ancient, and possibly a descendant of the bright blue sources only recently detected at $z>10$, or a more massive dust-obscured population. Not only will it serve as an important benchmark for future studies of high-redshift galaxy formation and evolution, the exquisite dataset associated with \galaxy\ makes it a prime test case for the next generation of sophisticated SED modeling codes. Future studies of \galaxy's morphology, environment, dynamics, and spatially resolved properties will offer additional insights into its puzzling birth and evolution. Uncovering the secrets of \galaxy's beginning will require continued excavation of this ancient ruin that formed at the dawn of time.

\begin{acknowledgments}

We gratefully acknowledge the JADES and SMILES teams for their survey planning and making their high-level science data products available to the astronomical community. Their efforts made our work possible. The data were obtained from the Mikulski Archive for Space Telescopes at the Space Telescope Science Institute, which is operated by the Association of Universities for Research in Astronomy, Inc., under NASA contract NAS 5-03127 for JWST. The data utilized are associated with programs $\#1180,\ \#1207$, $\#1895$, and $\#1963$. The comparison spectra in Figure~\ref{fig:progenitor_spec} were sourced from programs $\#1210$, $\#1287$,  $\#2565$, $\#4233$. The JADES photometric catalog GOODS-S-Deep-2.0 can be accessed via doi:10.17909/8tdj-8n28 (with data from JEMS; doi:10.17909/fsc4-dt61) and the SMILES photometric catalog hlsp$\_$smiles$\_$jwst$\_$miri$\_$goodss$\_$multi$\_$v1.0$\_$catalog.fits is available at doi:10.17909/et3f-zd57. Some of the data products presented herein were retrieved from the Dawn JWST Archive \citep[DJA][]{brammer_2023_8319596,Heintz2024, DeGraaff2025}. DJA is an initiative of the Cosmic Dawn Center (DAWN), which is funded by the Danish National Research Foundation under grant DNRF140.

We thank Joel Leja, Bingjie Wang, and David Setton for discussions about \texttt{Prospector} modeling and Aliza Beverage for discussions about \texttt{alf$\alpha$}. BF acknowledges support from JWST-GO-02913.001-A. GW gratefully acknowledges support from the National Science Foundation through grant AST-2347348. While writing this paper we made extensive use of the Astrophysics Data Service (ADS) and \texttt{arXiv} preprint repository.

\end{acknowledgments}

\facilities{JWST, HST}

\software{Matplotlib \citep{2007CSE.....9...90H}, 
        NumPy \citep{2020Natur.585..357H},
        astropy \citep{2013A&A...558A..33A,2018AJ....156..123A,2022ApJ...935..167A},
        Prospector \citep{Johnson2017, Johnson2021}, 
        dynesty \citep{Speagle2020}, 
        Bagpipes \citep{Carnall2018, Carnall2019}, 
        Nautilus \citep{Nautilus}, 
        SpectRes \citep{Carnall2017}
        alf$\alpha$ \citep{Beverage2025}, 
          mpi4py
          \citep{Dalcin2005, Dalcin2008, Dalcin2011, Dalcin2021, Rogowski2023}
}

\bibliography{bibby}{}
\bibliographystyle{aasjournal}

\appendix

\section{Effects of Repeatedly Sampling The Likelihood Surface}
\label{sec:sampling}

The Bayesian inference process used by \texttt{dynesty} identifies likelihood modes in the \texttt{Prospector} model posterior by dropping multiple random walkers down in the multi-dimensional parameter space to explore the likelihood surface and discover the maxima. However, as \citet{Wang2025} point out in their Appendix C, the likelihood space for SED models is very often complex and multimodal, especially for models with high dimensionality (such as the ones we use in this work). Therefore, out of an abundance of caution, we fit our observed data with our \texttt{Prospector} models multiple times to ensure that we do not infer properties from a low-likelihood local maximum. For our fiducial (alternative) model(s), we run the fits ten (five) times. Like \citet{Wang2025}, we also see that the sampler often identifies different local maxima in independent runs.

We experimented with using different batch sizes of random walkers in the exploration phase during which the likelihood surface is explored by walkers to search for the global maximum. Starting with exploration batch sizes of 2000 walkers (common for science applications; J. Leja, B. Wang priv. comm.), we increase the batch size by factors of two to 4000, 8000, 16,000, and 32,000. Using higher batch sizes should more completely sample the likelihood surface and we find that it does often infer better agreement in the inferred posteriors across independent runs, but we see diminishing returns. Even repeat fits with CPU-expensive 32,000 walker batches ($\sim12$ hours for a fit parallelized across 16 CPU cores) select different maxima. All of the fits presented in this work utilize an exploration phase with 8000 walkers per batch, fewer than the highest batch size tested but still erring on the side of excess compared to typical analyses. Using random slice sampling or multivariate slice sampling instead of random walkers does not seem to produce better agreement in inferred posteriors between independent runs. We find that the sampler also identifies false modes in nuisance parameters and some of these quantities' inferred posteriors which should be flat are instead peaked. This effect can be lessened by reducing the dimensionality of the model (for instance, removing nuisance parameters or choosing fewer SFH bins), but this is still troubling trend for future analyses of multiwavelength datasets of complex astrophysical sources which may require simultaneously modeling \emph{many} parameters e.g., stellar populations, dust emission, and AGN activity.

\begin{figure*}[!htb]
\centering
\includegraphics[width=\linewidth]{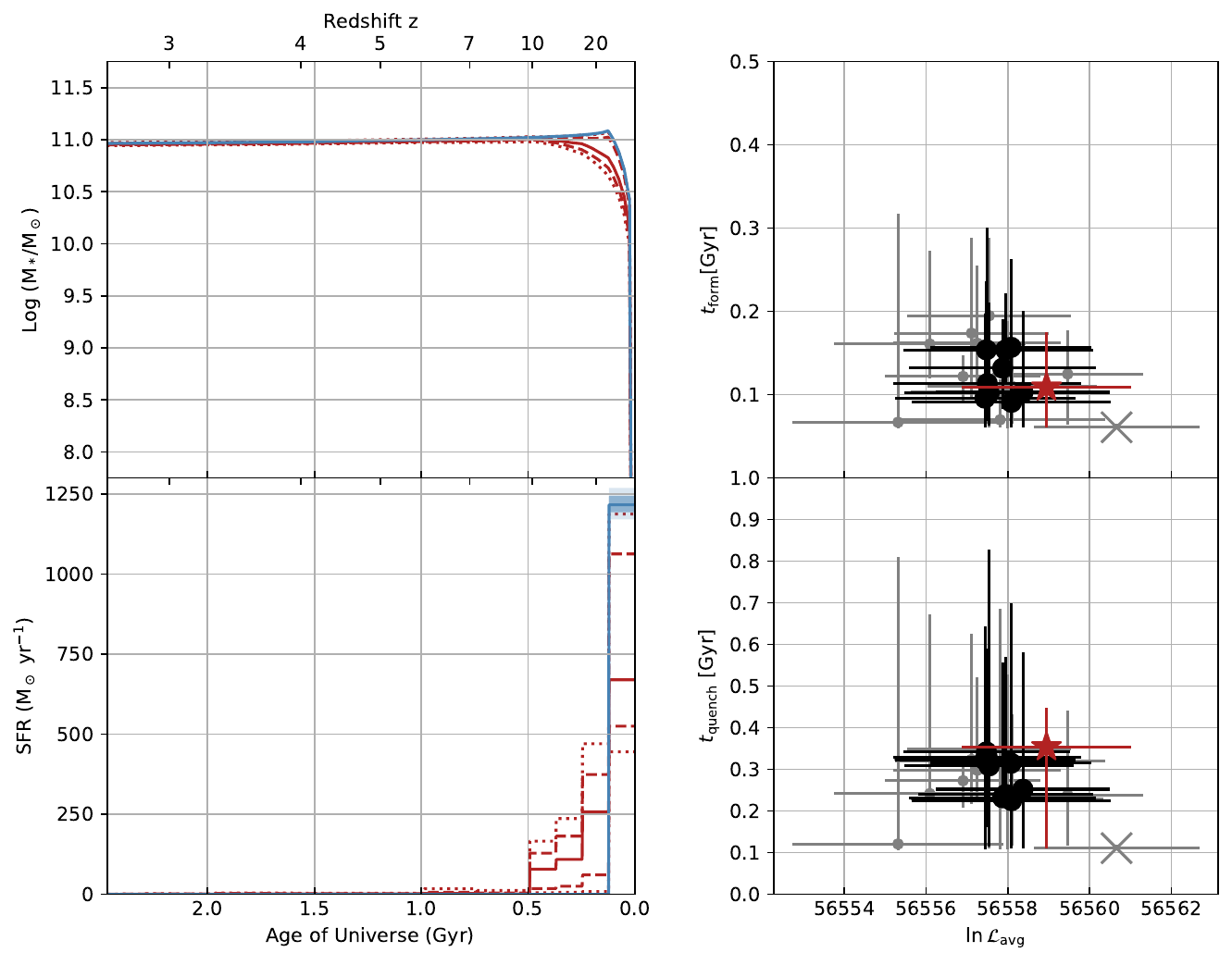}
\caption{\textbf{Left}: In blue we show the SFH posteriors inferred from the ``maximally early'' likelihood mode in one of our \texttt{Prospector} fits to the rising SFH model, similar to Figure~\ref{fig:sfh_alternative}, though we show SFR in linear scale to highlight the extremity of this mode. The dark/light shaded regions in both panels indicate $1\sigma$/$2\sigma$ uncertainties (yes, the uncertainty is included in the stellar mass panel too). We also show the fiducial model's median/$1\sigma$/$2\sigma$ histories by solid/dashed/dotted red lines. \textbf{Right:} Inferred $t_\mathrm{form}$ (top) and $t_\mathrm{quench}$ (bottom) versus weighted average likelihood $\log \mathcal{L}$ for our \texttt{Prospector} runs using batch sizes of 4000 (gray) and 8000 (black) random walkers in the exploration phase of the nested sampling. The run used in the analyses presented in this paper as the fiducial model is chosen to be the highest likelihood 8000 walker fit and is indicated by a red star. The gray X in the bottom right hand corner is a 4000 walker fit that found the maximally early mode, like the rising SFH fit shown in the left panels. The horizontal error bars are the weighted standard deviation of $\log \mathcal{L}$ and the vertical error bars show the $2\sigma$ uncertainty to highlight how small the posterior of the maximally old mode is in this parameter space. Clearly increasing the batch size of walkers in the exploration phase produces better agreement between fits, though we still find scatter even for very large and computationally expensive exploration batch sizes (i.e., 32,000 walkers).}
\label{fig:sampling}
\end{figure*}

One troubling maximum which we suspect exists in all of our models' parameter spaces (except perhaps the fixed metallicity model) is the maximally early formation mode. Occasionally during the exploration phase, the sampler identifies a location in parameter space which puts \emph{all} of the star formation in the very first bin with extremely small uncertainties. This maximum is identified when fit to the fiducial model once each in runs with 32,000, 16,000, and 4000 exploration walkers. It also appears with alternative models: once in a $z_{\mathrm{SF}}<10$ run and twice in the rising SFH model runs (one of which is shown in blue the left two panels of Figure~\ref{fig:sampling}). We have not identified a higher likelihood mode in any of our other runs, which would imply that this is mode is the global maximum. We remove these runs from consideration because the SFH uncertainties are so underestimated, even though they represent the highest likelihoods sampled in parameter space and generally agree with results from SSP tests which prefer $t_\mathrm{form}\approx0$ (see Appendix~\ref{sec:ssp}). Including these in our analysis would not qualitatively affect our results, it would primarily result in even earlier inferred $t_\mathrm{form}$ and $t_\mathrm{quench}$ times.

In the right two panels of Figure~\ref{fig:sampling} we show the weighted mean likelihoods and inferred $t_\mathrm{form}$ (upper) and $t_\mathrm{quench}$ (lower) values for the ten 8000 walker runs and ten 4000 walker runs, plus the one rejected 4000 walker run which fit the maximally early solution (gray cross in the bottom right). The 8000 walker run used as the ``fiducial model'' posterior in our analysis is highlighted as a red star. The vertical error bars show the $2\sigma$ uncertainties (2.5th and 97.5th percentiles) in $t_\mathrm{form}$ and $t_\mathrm{quench}$ to demonstrate the extremity of this maximally early solution. There appears to be an underlying trend that higher likelihood modes correspond with earlier formation and quenching (as seen when comparing different models as well), but our fits all produce very early formation times which reinforce our conclusions that \galaxy\ is old and formed early. We see the most variation in inferred SFHs in models which put priors on the SFH (i.e., rising SFH and Prospector-$\beta$) but these correspond with larger variation in weighted average likelihoods for the fits. Even though these models \emph{can} infer later $t_\mathrm{form}$/$t_\mathrm{quench}$, these solutions are always associated with significantly lower likelihood modes and are disfavored. That the \texttt{dynesty} sampling gets ``stuck'' in or does not identify certain modes implies that the uncertainties in SFHs inferred with \texttt{Prospector} in general may be significantly underestimated. 


\section{SSP Tests}
\label{sec:ssp}

To acquire a lower limit on the estimated age of \galaxy, we fit it with an SSP \citep{Conroy2013, Carnall2024}. This is equivalent to a SFH consisting of a single instantaneous burst of star formation and its simplicity will allow for a more direct comparison between \texttt{Prospector} and \texttt{Bagpipes} models. We will start with \texttt{Bagpipes}.

SFH models in \texttt{Bagpipes} (including the burst model) require that no stars can form before the Big Bang. Therefore, the age of the SSP fit with \texttt{Bagpipes} is younger than the age of the universe by definition. We apply a uniform prior on age $t_\mathrm{age}\in[0, t_\mathrm{obs}]$.

We perform two separate SSP fits with \texttt{Prospector}. In the first fit, Prospector$^{a}$, we require that the age of the SSP is less than the age of the universe at $z_\mathrm{obs}$ with a flat prior (i.e., as the \texttt{Bagpipes} fit). In the second fit, Prospector$^{b}$, we place no restriction on the SSP's age.

Our results are shown in the top half of Table~\ref{tab:ssp}. In \texttt{Bagpipes}, the single burst of star formation produces a formation time consistent with the double power law, indicating that \galaxy\ formed by $z_\mathrm{burst}>8$ at the latest. Our \texttt{Prospector} SSP fits, however, infer $t_\mathrm{form}\approx0$ ($t_\mathrm{age}\approx t_\mathrm{obs}$). Without any restriction to age (Prospector$^b$), we find that \texttt{Prospector} prefers an SSP which is slightly older than the age of the universe.

We speculate that this discrepancy between SSP ages inferred with \texttt{Bagpipes} and \texttt{Prospector} is driven by differences in stellar libraries and isochrones. Typically as stellar ages increase, the ability to observationally resolve difference in stellar ages decreases \citep[e.g.,][]{Conroy2013}. A log age difference of 0.1 dex between SSPs fit with different SED modeling codes is not catastrophic, though this does demonstrate that systematic uncertainties play a significant role and care must be taken in interpreting results as $t_\mathrm{age}$ approaches $t_\mathrm{obs}$. Other confounding variables which could affect the differing SSP ages include under-the-hood differences in modeling, such as likelihood functions, nested sampling implementations \citep[see Appendix \ref{sec:sampling} and][]{Wang2025}, or how the spectroscopy and photometry is weighted in the fits (as might be suggested by the slight difference in inferred $z_\mathrm{spec}$ seen in Table~\ref{tab:pipetable}).

\begin{deluxetable*}{ccccc}
\centerwidetable
\tablewidth{0pt}
\tablecaption{Top: Stellar mass and formation times for \galaxy\ from inferred from different SSP fits. Bottom: inferred SSP properties of \galaxy\ compared with ZF-UDS-7329 and PRIMER-EXCELS-109760 from \citet{Carnall2024}.\label{tab:ssp}}
\tablehead{\colhead{Fit} & \colhead{$\log M_*/\mathrm{M_\odot}$} & \colhead{$t_\mathrm{age}$} & \colhead{$t_\mathrm{form}$} & \colhead{$z_\mathrm{form}$} \\
\colhead{} & \colhead{} & \colhead{[Gyr]} & \colhead{[Myr]} & \colhead{}
}
\startdata
Bagpipes &	$10.92\pm0.01$	&	$1.99\pm0.16$ & $480^{+158}_{-164}$	&	$9.98_{-1.90}^{+3.53}$\\
Prospector$^a$ & $10.96\pm0.01$& $2.38^{+0.06}_{-0.09}$ & $82^{+89}_{-60}$ & $33.61_{-13.06}^{+42.13}$\\
Prospector$^b$ & $10.98\pm0.02$ & $2.51^{+0.18}_{-0.15}$ & $-42_{-177}^{+152}$ & $>28.25^{c}$\\
\hline
\hline
\colhead{ID} & \colhead{$\log M_*/\mathrm{M_\odot}$} & \colhead{$t_\mathrm{age}$} & \colhead{$t_\mathrm{form}$} & \colhead{$z_\mathrm{form}$} \\
\colhead{} & \colhead{} & \colhead{[Gyr]} & \colhead{[Myr]} & \colhead{}\\
\hline
\galaxylong & $10.90\pm0.01$ & $1.99\pm0.16$  & $402^{+164}_{-163}$ & $11.14^{+5.03}_{-2.48}$\\
PRIMER-EXCELS-109760$^d$ & $11.01\pm0.03$ & --- & $550\pm110$ & $8.8^{+1.6}_{-1.1}$\\
ZF-UDS-7329$^d$ & $11.12^{+0.04}_{-0.07}$ & --- & $490^{+240}_{-220}$ & $9.6^{+5.2}_{-2.4}$
\enddata\
\tablenotetext{a}{Requiring $t_\mathrm{age} < t_\mathrm{obs}$ (i.e., $t_\mathrm{form} > 0$)}
\tablenotetext{b}{Any $t_\mathrm{age}$ allowed}
\tablenotetext{c}{We use the upper $1\sigma$ uncertainty on $t_\mathrm{form}$ as the lower limit for $z_\mathrm{form}$}
\tablenotetext{d}{Values from \citet{Carnall2024}}
\end{deluxetable*}

\subsection{Comparison with EXCELS}

To directly compare \galaxy\ with the SSP ages of ZF-UDS-7329 and PRIMER-EXCELS-109760 in \citet{Carnall2024}, we perform an additional SSP fit with \texttt{Bagpipes} using a matching flat $\Lambda$CDM cosmology: $H_0 = 70 \ \mathrm{km \ s^{-1} \ Mpc^{-1}}$, $\Omega_m = 0.30$, and $\Omega_\Lambda = 0.70$. We show our results in Table~\ref{tab:ssp}. Unsurprisingly, changing the cosmology has virtually no effects on the age of the SSP, though the new cosmology pushes $t_\mathrm{form}$/$z_\mathrm{form}$ even earlier due to the the difference in inferred $\Delta t_\mathrm{obs}\approx75\ \mathrm{Myr}$. We find that SSP formation time for \galaxy\ ($z_\mathrm{form}=11.14^{+5.03}_{-2.48}$) is very similar to those inferred for ZF-UDS-7329 ($z_\mathrm{burst}=9.4^{+5.2}_{-2.4}$) and PRIMER-EXCELS-109760 ($z_\mathrm{burst}=8.8^{+1.6}_{-1.1}$) in \citet{Carnall2024}, which suggests that these galaxies could share common ancestors.

\begin{rotatetable*}
\begin{deluxetable*}{cccccccc}
\centerwidetable
\tablewidth{0pt}
\tablecaption{Formation of \galaxy\ and Fit Statistics Inferred by SED Modeling \label{tab:megatable2}}
\tablehead{\colhead{Model} & \colhead{$t_\mathrm{form}$} & \colhead{$z_\mathrm{form}$} & \colhead{$t_\mathrm{quench}$} & \colhead{$z_\mathrm{quench}$} & \colhead{$\chi^{2}_{\mathrm{phot}}$} & \colhead{$\chi^{2}_{\mathrm{spec}}$} & \colhead{$\ln\ \mathcal{L}_\mathrm{avg}$ ($\ln\ \mathcal{L}_\mathrm{max}$)}\\
\colhead{} & \colhead{$\mathrm{[Myr]}$} & \colhead{} & \colhead{$\mathrm{[Myr]}$} & \colhead{} & \colhead{} & \colhead{} & \colhead{}}
\startdata
Fiducial	&	$108_{-39}^{+37}\ (69)$	&	$28.41_{-5.29}^{+10.31}\ (38.40)$	&	$353_{-179}^{+53}\ (207)$	&	$12.46_{-1.20}^{+8.10}\ (18.15)$	&	35.00	&	4461.35	&	$56558.95\pm2.03\ (56564.12)$\\
\hline
Without MIRI	&	$88_{-21}^{+25}\ (81)$	&	$32.85_{-5.19}^{+6.72}\ (34.74)$	&	$218_{-98}^{+65}\ (202)$	&	$17.55_{-2.98}^{+8.99}\ (18.52)$	&	36.38$^a$	&	4460.86	&	$56557.12\pm2.30\ (56563.67)$$^a$\\
G140M	&	$213_{-52}^{+62}\ (144)$	&	$17.83_{-2.96}^{+3.85}\ (23.41)$	&	$534_{-137}^{+145}\ (465)$	&	$9.23_{-1.51}^{+2.24}\ (10.22)$	&	35.76	&	2418.89$^a$	&	$29936.30\pm2.11\ (29941.59)$$^a$\\
G235M	&	$125_{-37}^{+145}\ (90)$	&	$25.80_{-10.74}^{+7.08}\ (32.33)$	&	$465_{-243}^{+184}\ (353)$	&	$10.21_{-2.24}^{+7.11}\ (12.48)$	&	32.12	&	2056.91$^a$	&	$29519.50\pm2.11\ (29524.19)$$^a$\\
\hline
$z_\mathrm{SF}<10$	&	$581_{-42}^{+41}\ (529)$	&	$8.67_{-0.43}^{+0.50}\ (9.29)$	&	$697_{-58}^{+75}\ (570)$	&	$7.56_{-0.57}^{+0.51}\ (8.79)$	&	115.20	&	4482.09	&	$56552.16\pm2.17\ (56558.07)$\\
$z_\mathrm{SF}<15$	&	$328_{-1}^{+18}\ (326)$	&	$13.15_{-0.51}^{+0.04}\ (13.18)$	&	$373_{-2}^{+104}\ (370)$	&	$11.99_{-1.97}^{+0.06}\ (12.04)$	&	39.45	&	4467.85	&	$56552.32\pm2.49\ (56558.64)$\\
$z_\mathrm{SF}<20$	&	$245_{-7}^{+26}\ (269)$	&	$16.13_{-1.14}^{+0.37}\ (15.10)$	&	$338_{-53}^{+85}\ (378)$	&	$12.87_{-1.94}^{+1.67}\ (11.86)$	&	37.33	&	4465.75	&	$56555.09\pm2.17\ (56560.75)$\\
\hline
Bursty SFH	&	$65_{-4}^{+38}\ (60)$	&	$39.97_{-10.67}^{+2.11}\ (42.14)$	&	$118_{-8}^{+134}\ (110)$	&	$26.73_{-10.93}^{+1.43}\ (28.04)$	&	35.87	&	4459.26	&	$56559.26\pm2.39\ (56564.73)$\\
Rising SFH	&	$129_{-43}^{+54}\ (72)$	&	$25.15_{-5.38}^{+7.97}\ (37.65)$	&	$345_{-99}^{+271}\ (178)$	&	$12.68_{-4.38}^{+3.45}\ (20.23)$	&	39.81	&	4464.14	&	$56555.56\pm2.13\ (56560.32)$\\
Prospector-$\beta$	&	$215_{-99}^{+120}\ (119)$	&	$17.68_{-4.75}^{+9.43}\ (26.60)$	&	$623_{-114}^{+61}\ (570)$	&	$8.23_{-0.56}^{+1.33}\ (8.79)$	&	54.61	&	4462.95	&	$56554.37\pm2.86\ (56562.39)$\\
\hline
Kroupa IMF	&	$88_{-21}^{+42}\ (62)$	&	$32.62_{-7.71}^{+6.61}\ (41.23)$	&	$306_{-179}^{+55}\ (113)$	&	$13.79_{-1.54}^{+11.78}\ (27.53)$	&	35.67	&	4459.47	&	$56558.96\pm2.18\ (56564.72)$\\
$\log(Z/\mathrm{Z_\odot})=0.19$	&	$659_{-187}^{+142}\ (254)$	&	$7.89_{-1.08}^{+2.21}\ (15.73)$	&	$1423_{-55}^{+33}\ (1438)$	&	$4.32_{-0.08}^{+0.14}\ (4.28)$	&	55.72	&	4476.39	&	$56540.93\pm2.02\ (56546.24)$\\
\enddata
\tablecomments{The inferred formation and quenching times/redshifts for \galaxy. We also show the average log likelihood ($\ln\ \mathcal{L}_\mathrm{avg}$) and the $\chi^2$ values for the MAP model for the spectrum and photometry individually. While the MAP models all have very similar $\chi^2_\mathrm{spec}$, later-forming models ($z_\mathrm{SF}<10$, Prospector-$\beta$, and $\log(Z/\mathrm{Z_\odot})=0.19$) all have higher $\chi^2_\mathrm{phot}$ and $\ln\ \mathcal{L}_\mathrm{avg}$. There is a noticeable trend that the MAP $t_\mathrm{form}$ and $t_\mathrm{quench}$ times are earlier than the inferred posterior distribution.}
\tablenotetext{a}{These quantities are given for completeness. These fits run with differing datasets and thus should not be used for comparison with other models.}
\tablecomments{\textcolor{blue}{Due to a bug in the implementation of the rotatetable environment within the text in \aastex\ v7 (fix pending), this table has been moved to the end of the document where it does not interfere with the text.}}
\end{deluxetable*}
\end{rotatetable*}

\end{document}